\documentclass[prd,showkeys,floatfix,twocolumn,amsmath,amssymb,floatfix]{revtex4}
\usepackage{graphicx}
\usepackage{multirow}
\usepackage{subfigure}
\usepackage[colorlinks,citecolor=blue,anchorcolor=red,menucolor=red,linkcolor=red,filecolor=red,runcolor=red,urlcolor=blue,frenchlinks=red]{hyperref}
\usepackage{enumitem}

\makeatletter
\renewcommand{\@thesubfigure}{\hskip\subfiglabelskip}
\makeatother
\allowdisplaybreaks[3]

\begin{document}
\title{Light double-gluon hybrid states with the exotic quantum numbers $J^{PC} = 1^{-+}$ and $3^{-+}$}
%

\author{Niu Su$^1$}
\author{Wei-Han Tan$^1$}
\author{Hua-Xing Chen$^1$}
\email{hxchen@seu.edu.cn}
\author{Wei Chen$^2$}
\email{chenwei29@mail.sysu.edu.cn}
\author{Shi-Lin Zhu$^3$}
\email{zhusl@pku.edu.cn}

\affiliation{
$^1$School of Physics, Southeast University, Nanjing 210094, China\\
$^2$School of Physics, Sun Yat-Sen University, Guangzhou 510275, China\\
$^3$School of Physics and Center of High Energy Physics, Peking University, Beijing 100871, China}
\begin{abstract}
We apply the QCD sum rule method to study the double-gluon hybrid states with the quark-gluon contents $\bar q q gg$ ($q=u/d$) and $\bar s s gg$. We construct twenty-eight double-gluon hybrid currents, eleven of which are found to be zero due to some internal symmetries between the two gluons fields. We concentrate on the non-vanishing currents with the exotic quantum numbers $J^{PC} = 1^{-+}$ and $3^{-+}$. Their masses are calculated to be $M_{|\bar q q gg;1^{-+}\rangle} = 4.35^{+0.26}_{-0.30}$~GeV, $M_{|\bar s s gg;1^{-+}\rangle} = 4.49^{+0.25}_{-0.30}$~GeV, $M_{|\bar q q gg;3^{-+}\rangle} = 3.02^{+0.24}_{-0.31}$~GeV, and $M_{|\bar s s gg;3^{-+}\rangle} = 3.16^{+0.22}_{-0.28}$~GeV. The decay behaviors of the $J^{PC} = 3^{-+}$ states are studied, and we propose to search for them in the $\pi a_1(1260)/\rho \omega/\phi \phi$ channels in future particle experiments.
\end{abstract}
\maketitle
\pagenumbering{arabic}

\section{Introduction}
\label{sec:intro}

A double-gluon hybrid state consists of one valence quark, one valence antiquark, and two valence gluons. Some of them have the exotic quantum numbers that the traditional $\bar q q$ mesons can not reach, {\it e.g.}, $J^{PC} =0^{--}/0^{+-}/1^{-+}/2^{+-}/3^{-+}/4^{+-}/\cdots$, and these states are of particular interests. Until now, there are four states observed in particle experiments, which have the exotic quantum number $J^{PC}=1^{-+}$~\cite{pdg}. They are the $\pi_1(1400)$~\cite{IHEP-Brussels-LosAlamos-AnnecyLAPP:1988iqi}, $\pi_1(1600)$~\cite{E852:1998mbq}, $\pi_1(2015)$~\cite{E852:2004gpn}, and $\eta_1(1855)$~\cite{BESIII:2022riz,BESIII:2022qzu}. The former three have $I=1$ and the latter one has $I=0$. These four structures are candidates for the single-gluon hybrid states, which have been extensively studied during the past fifty years, both experimentally and theoretically~\cite{Barnes:1977hg,Hasenfratz:1980jv,Chanowitz:1982qj,Isgur:1983wj,Close:1994hc,Page:1998gz,Burns:2006wz,Qiu:2022ktc,Szczepaniak:2001rg,Iddir:2007dq,Guo:2007sm,Michael:1985ne,McNeile:1998cp,Juge:2002br,Lacock:1996ny,MILC:1997usn,Bernard:2003jd,Hedditch:2005zf,Dudek:2009qf,Dudek:2010wm,Dudek:2013yja,Chen:2022isv,Balitsky:1982ps,Govaerts:1983ka,Kisslinger:1995yw,Chetyrkin:2000tj,Jin:2002rw,Narison:2009vj,Huang:2016upt,Li:2021fwk,COMPASS:2009xrl}. However, their nature is still elusive due to our poor understanding of the gluon degree of freedom~\cite{Chen:2022asf,Klempt:2007cp,Meyer:2015eta,Amsler:2004ps,Bugg:2004xu,Meyer:2010ku,Briceno:2017max,COMPASS:2018uzl,JPAC:2018zyd,Ketzer:2019wmd,Jin:2021vct,Meng:2022ozq,Albuquerque:2023bex}. Besides, these states can also be interpreted as the compact tetraquarks and hadronic molecules~\cite{Chen:2008qw,Chen:2008ne,Zhang:2019ykd,Dong:2022cuw,Yang:2022lwq,Wan:2022xkx,Wang:2022sib,Su:2022eun,Yu:2022wtu}.

Recently, the D0 and TOTEM collaborations compared their $pp$ and $p\bar p$ cross sections, and observed the evidence of a $C$-odd three-gluon glueball~\cite{D0:2012erd,COMPETE:2002jcr,TOTEM:2017sdy,Martynov:2018sga,TOTEM:2020zzr}. Inspired by these experiments, we studied the double-gluon hybrid states in Refs.~\cite{Chen:2021smz,Su:2022fqr} based on our previous studies on the single-gluon hybrid states~\cite{Huang:2010dc,Chen:2010ic,Chen:2022qpd}. In these studies we applied the QCD sum rule method to investigate some double-gluon hybrid states, and we mainly concentrated on the states with the exotic quantum number $J^{PC}=2^{+-}$. More QCD sum rule studies can be found in Ref.~\cite{Tang:2021zti}, where the authors investigated the double-gluon charmonium and bottomonium hybrid states of $J^{PC}=1^{--}$.

In this paper we shall investigate more double-gluon hybrid states through the QCD sum rule method. Especially, we shall concentrate on the states with the exotic quantum numbers $J^{PC}=1^{-+}$ and $3^{-+}$. Assuming their quark-gluon contents to be either $\bar q q gg$ ($q=u/d$) or $\bar s s gg$, we shall calculate their masses to be:
\begin{eqnarray}
\nonumber M_{|\bar q q gg;1^{-+}\rangle} &=& 4.35^{+0.26}_{-0.30}{\rm~GeV} \, ,
\\ \nonumber M_{|\bar s s gg;1^{-+}\rangle} &=& 4.49^{+0.25}_{-0.30}{\rm~GeV} \, ,
\\ \nonumber M_{|\bar q q gg;3^{-+}\rangle} &=& 3.02^{+0.24}_{-0.31}{\rm~GeV} \, ,
\\ \nonumber M_{|\bar s s gg;3^{-+}\rangle} &=& 3.16^{+0.22}_{-0.28}{\rm~GeV} \, .
\end{eqnarray}
Especially, the masses of the $J^{PC} = 3^{-+}$ states are calculated to be slightly larger than 3.0~GeV, which values are accessible at BESIII, Belle-II, GlueX, LHCb, and PANDA, etc. We shall also study their decay behaviors in this study.

This paper is organized as follows. In Sec.~\ref{sec:current} we construct twenty-eight double-gluon hybrid currents, eleven of which are found to be zero due to some internal symmetries between the two gluon fields. We apply the QCD sum rule method to study these currents in Sec.~\ref{sec:sumrule}, and then we perform numerical analyses in Sec.~\ref{sec:numerical}. The obtained results are summarized and discussed in Sec.~\ref{sec:summary}.

\section{Double-gluon hybrid currents}
\label{sec:current}

In this section we construct the double-gluon hybrid currents using the light quark field $q_a(x)$ and the gluon field strength tensor $G^n_{\mu\nu}(x)$. Besides, we also need the light antiquark field $\bar q_a(x)$ and the dual gluon field strength tensor $\tilde G^n_{\mu\nu} = G^{n,\rho\sigma} \times \epsilon_{\mu\nu\rho\sigma}/2$. Here $a=1\cdots3$ and $n=1\cdots8$ are color indices, and $\mu,\nu,\rho,\sigma$ are Lorentz indices.

Generally speaking, we can construct the double-gluon hybrid currents by combining the color-octet quark-antiquark fields
\begin{eqnarray}
\nonumber &\bar q_a \lambda_n^{ab} q_b \, , \, \bar q_a \lambda_n^{ab} \gamma_5 q_b \, , \, &
\\ & \bar q_a \lambda_n^{ab} \gamma_\mu q_b \, , \, \bar q_a \lambda_n^{ab} \gamma_\mu \gamma_5 q_b \, , \, &
\\ \nonumber & \bar q_a \lambda_n^{ab} \sigma_{\mu\nu} q_b \, , &
\end{eqnarray}
and the color-octet double-gluon fields
\begin{equation}
d^{npq} G_p^{\alpha\beta} G_q^{\gamma\delta} \, , \, f^{npq} G_p^{\alpha\beta} G_q^{\gamma\delta} \, ,
\end{equation}
together with some Lorentz coefficients $\Gamma^{\mu\nu\cdots\alpha\beta\gamma\delta}$. Here $d^{npq}$ and $f^{npq}$ are the totally symmetric and antisymmetric $SU(3)$ structure constants, respectively. Some other double-gluon hybrid currents can be constructed by combining color-singlet quark-antiquark fields and color-singlet double-gluon fields, {\it i.e.},
\begin{equation}
\bar q_a q_a \otimes G_n G_n \, .
\end{equation}
However, these currents are not strongly correlated, so we shall not investigate them in the present study.

Besides the double-gluon components, the double-gluon hybrid currents also contain the triple-gluon and quadruple-gluon components, because the gluon field strength tensor $G^n_{\mu\nu}$ is defined as
\begin{equation}
G^n_{\mu\nu} = \partial_\mu A_\nu^n  -  \partial_\nu A_\mu^n  +  g_s f^{npq} A_{p,\mu} A_{q,\nu} \, .
\end{equation}
Therefore, there can exist a significant mixing among the single-/double-/triple-/quadruple-gluon hybrid states of various color configurations, and moreover, these states can also mix with the conventional mesons, tetraquark states, and double-/triple-/quadruple-gluon glueballs, etc. We shall not investigate this effect in the present study, and note that there is still a long long way to understand glueballs and hybrid states as well as the gluon degree of freedom.

The double-gluon hybrid currents with the color-octet quark-antiquark field $\bar q_a \lambda_n^{ab} \gamma_5 q_b$ have been systematically constructed in Refs.~\cite{Chen:2021smz,Su:2022fqr}. These currents have the spin-parity quantum numbers $J^{PC} = 0^{\pm+}/1^{\pm-}/2^{\pm+}/2^{+-}$. Especially, the masses of the states with the exotic quantum number $J^{PC} = 2^{+-}$ are calculated to be
\begin{eqnarray}
M_{|\bar q q gg;2^{+-}\rangle} &=& 2.26^{+0.20}_{-0.25} {\rm~GeV} \, ,
\\
M_{|\bar s s gg;2^{+-}\rangle} &=& 2.38^{+0.19}_{-0.25} {\rm~GeV} \, .
\end{eqnarray}
However, these currents can not reach the quantum numbers $J^{PC} =0^{\pm-}/1^{\pm+}/2^{--}/3^{\pm\pm}/\cdots $, and especially, they can not reach the exotic quantum numbers $J^{PC} = 1^{-+}$ and $3^{-+}$.

\begin{figure*}[hbt]
\begin{center}
\includegraphics[width=0.75\textwidth]{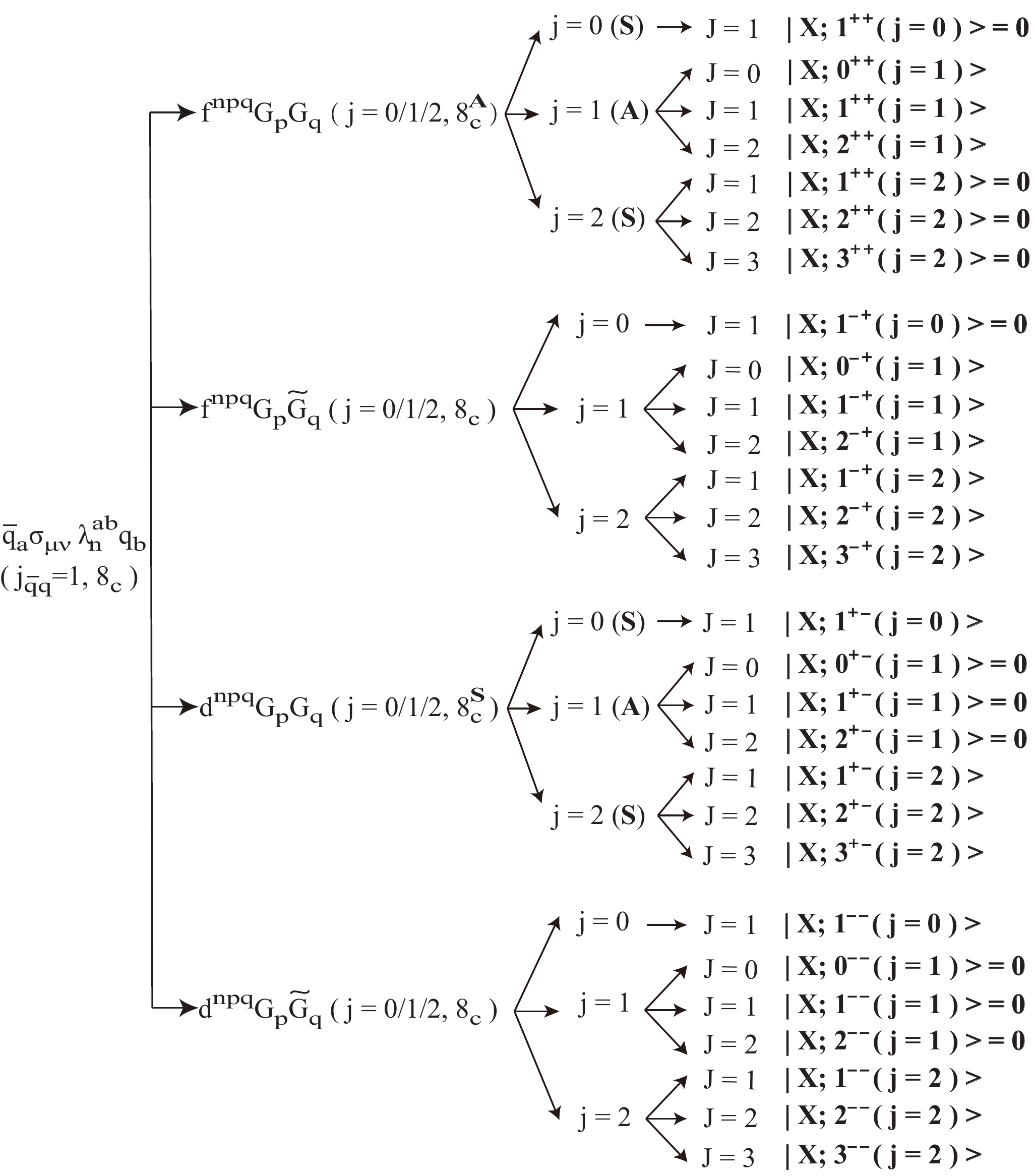}
\caption{Categorization of the double-gluon hybrid currents containing the color-octet quark-antiquark field $\bar q_a \lambda_n^{ab} \sigma_{\mu\nu} q_b$, where $\mathbf{A}$ and $\mathbf{S}$ denote the structure to be antisymmetric and symmetric, respectively.}
\label{fig:category}
\end{center}
\end{figure*}

In the present study we further construct the double-gluon hybrid currents using the color-octet quark-antiquark field $\bar q_a \lambda_n^{ab} \sigma_{\mu\nu} q_b$. We denote them as $J^{\alpha_1\beta_1\cdots\alpha_J\beta_J}_{J^{PC}(j)}$, where $J$ is the total spin of the double-gluon hybrid current, and $j$ is the spin of the double-gluon field. As summarized in Fig.~\ref{fig:category}, there are altogether twenty-eight currents:
\begin{widetext}
\begin{eqnarray}
J^{\alpha\beta}_{1^{++}(j=0)} &=& \bar q_a \sigma^{\alpha\beta} \lambda_n^{ab} q_b~f^{npq}~g_s^2 G_p^{\mu\nu} G_{q,\mu\nu} \, ,
\label{def:01pp}
\\
J_{0^{++}(j=1)} &=& \bar q_a \sigma^{\mu\nu} \lambda_n^{ab} q_b~f^{npq}~g_s^2 G_{p,\nu\rho} G_{q,\mu}^\rho \, ,
\label{def:10pp}
\\
J^{\alpha\beta}_{1^{++}(j=1)} &=& \bar q_a \sigma_{\alpha_1\beta_1} \lambda_n^{ab} q_b~f^{npq}~g_s^2 \left( G_{p,\alpha_2\mu} G_{q,\beta_2}^{\mu} - \{ \alpha_2 \leftrightarrow \beta_2 \} \right) \times g^{\beta_1 \beta_2} (g^{\alpha\alpha_1}g^{\beta\alpha_2} - g^{\beta\alpha_1}g^{\alpha\alpha_2} ) \, ,
\label{def:11pp}
\\
J^{\alpha_1\beta_1,\alpha_2\beta_2}_{2^{++}(j=1)} &=& \mathcal{S}[ \bar q_a \sigma^{\alpha_1\beta_1} \lambda_n^{ab} q_b~f^{npq}~g_s^2 \left( G_p^{\alpha_2\mu} G_{q,\mu}^{\beta_2} - \{ \alpha_2 \leftrightarrow \beta_2 \} \right) ] \, ,
\label{def:12pp}
\\
J^{\alpha\beta}_{1^{++}(j=2)} &=& \bar q_a \sigma^{\mu_2\nu_2} \lambda_n^{ab} q_b~f^{npq}~\mathcal{S}^\prime[ g_s^2 G_{p,\mu_1\nu_1} G_{q,\mu_2\nu_2}] \times g^{\alpha\mu_1}g^{\beta\nu_1} \, ,
\label{def:21pp}
\\
J^{\alpha_1\beta_1,\alpha_2\beta_2}_{2^{++}(j=2)} &=&  \mathcal{S}[ \bar q_a \sigma_{\mu_3\nu_3} \lambda_n^{ab} q_b~f^{npq}~\mathcal{S}^\prime[ g_s^2 G_{p,\mu_1\nu_1} G_{q,\mu_2\nu_2}] \times g^{\alpha_1\mu_1}g^{\beta_1\nu_1}g^{\nu_2\nu_3}(g^{\alpha_2\mu_2}g^{\beta_2\mu_3} - g^{\beta_2\mu_2}g^{\alpha_2\mu_3})] \, ,
\label{def:22pp}
\\
J^{\alpha_1\beta_1,\alpha_2\beta_2,\alpha_3\beta_3}_{3^{++}(j=2)} &=& \mathcal{S}[ \bar q_a \sigma^{\alpha_1\beta_1} \lambda_n^{ab} q_b~f^{npq}~g_s^2 G_p^{\alpha_2\beta_2} G_q^{\alpha_3\beta_3}] \, ,
\label{def:23pp}
\\ 
J^{\alpha\beta}_{1^{-+}(j=0)} &=& \bar q_a \sigma^{\alpha\beta} \lambda_n^{ab} q_b~f^{npq}~g_s^2 G_p^{\mu\nu} \tilde G_{q,\mu\nu} \, ,
\label{def:01mp}
\\
J_{0^{-+}(j=1)} &=& \bar q_a \sigma^{\mu\nu} \lambda_n^{ab} q_b~f^{npq}~g_s^2 G_{p,\nu\rho} \tilde G_{q,\mu}^\rho \, ,
\label{def:10mp}
\\
J^{\alpha\beta}_{1^{-+}(j=1)} &=& \bar q_a \sigma_{\alpha_1\beta_1} \lambda_n^{ab} q_b~f^{npq}~g_s^2 \left( G_{p,\alpha_2\mu} \tilde G_{q,\beta_2}^{\mu} - \{ \alpha_2 \leftrightarrow \beta_2 \} \right) \times g^{\beta_1 \beta_2} (g^{\alpha\alpha_1}g^{\beta\alpha_2} - g^{\beta\alpha_1}g^{\alpha\alpha_2} ) \, ,
\label{def:11mp}
\\
J^{\alpha_1\beta_1,\alpha_2\beta_2}_{2^{-+}(j=1)} &=& \mathcal{S}[ \bar q_a \sigma^{\alpha_1\beta_1} \lambda_n^{ab} q_b~f^{npq}~g_s^2 \left( G_p^{\alpha_2\mu} \tilde G_{q,\mu}^{\beta_2} - \{ \alpha_2 \leftrightarrow \beta_2 \} \right) ] \, ,
\label{def:12mp}
\\
J^{\alpha\beta}_{1^{-+}(j=2)} &=& \bar q_a \sigma^{\mu_2\nu_2} \lambda_n^{ab} q_b~f^{npq}~\mathcal{S}^\prime[ g_s^2 G_{p,\mu_1\nu_1} \tilde G_{q,\mu_2\nu_2}] \times g^{\alpha\mu_1}g^{\beta\nu_1} \, ,
\label{def:21mp}
\\
J^{\alpha_1\beta_1,\alpha_2\beta_2}_{2^{-+}(j=2)} &=&  \mathcal{S}[ \bar q_a \sigma_{\mu_3\nu_3} \lambda_n^{ab} q_b~f^{npq}~\mathcal{S}^\prime[ g_s^2 G_{p,\mu_1\nu_1} \tilde G_{q,\mu_2\nu_2}] \times g^{\alpha_1\mu_1}g^{\beta_1\nu_1}g^{\nu_2\nu_3}(g^{\alpha_2\mu_2}g^{\beta_2\mu_3} - g^{\beta_2\mu_2}g^{\alpha_2\mu_3})] \, ,
\label{def:22mp}
\\
J^{\alpha_1\beta_1,\alpha_2\beta_2,\alpha_3\beta_3}_{3^{-+}(j=2)} &=& \mathcal{S}[ \bar q_a \sigma^{\alpha_1\beta_1} \lambda_n^{ab} q_b~f^{npq}~g_s^2 G_p^{\alpha_2\beta_2} \tilde G_q^{\alpha_3\beta_3}] \, ,
\label{def:23mp}
\\ 
J^{\alpha\beta}_{1^{+-}(j=0)} &=& \bar q_a \sigma^{\alpha\beta} \lambda_n^{ab} q_b~d^{npq}~g_s^2 G_p^{\mu\nu} G_{q,\mu\nu} \, ,
\label{def:01pm}
\\
J_{0^{+-}(j=1)} &=& \bar q_a \sigma^{\mu\nu} \lambda_n^{ab} q_b~d^{npq}~g_s^2 G_{p,\nu\rho} G_{q,\mu}^\rho \, ,
\label{def:10pm}
\\
J^{\alpha\beta}_{1^{+-}(j=1)} &=& \bar q_a \sigma_{\alpha_1\beta_1} \lambda_n^{ab} q_b~d^{npq}~g_s^2 \left( G_{p,\alpha_2\mu} G_{q,\beta_2}^{\mu} - \{ \alpha_2 \leftrightarrow \beta_2 \} \right) \times g^{\beta_1 \beta_2} (g^{\alpha\alpha_1}g^{\beta\alpha_2} - g^{\beta\alpha_1}g^{\alpha\alpha_2} ) \, ,
\label{def:11pm}
\\
J^{\alpha_1\beta_1,\alpha_2\beta_2}_{2^{+-}(j=1)} &=& \mathcal{S}[ \bar q_a \sigma^{\alpha_1\beta_1} \lambda_n^{ab} q_b~d^{npq}~g_s^2 \left( G_p^{\alpha_2\mu} G_{q,\mu}^{\beta_2} - \{ \alpha_2 \leftrightarrow \beta_2 \} \right) ] \, ,
\label{def:12pm}
\\
J^{\alpha\beta}_{1^{+-}(j=2)} &=& \bar q_a \sigma^{\mu_2\nu_2} \lambda_n^{ab} q_b~d^{npq}~\mathcal{S}^\prime[ g_s^2 G_{p,\mu_1\nu_1} G_{q,\mu_2\nu_2}] \times g^{\alpha\mu_1}g^{\beta\nu_1} \, ,
\label{def:21pm}
\\
J^{\alpha_1\beta_1,\alpha_2\beta_2}_{2^{+-}(j=2)} &=&  \mathcal{S}[ \bar q_a \sigma_{\mu_3\nu_3} \lambda_n^{ab} q_b~d^{npq}~\mathcal{S}^\prime[ g_s^2 G_{p,\mu_1\nu_1} G_{q,\mu_2\nu_2}] \times g^{\alpha_1\mu_1}g^{\beta_1\nu_1}g^{\nu_2\nu_3}(g^{\alpha_2\mu_2}g^{\beta_2\mu_3} - g^{\beta_2\mu_2}g^{\alpha_2\mu_3})] \, ,
\label{def:22pm}
\\
J^{\alpha_1\beta_1,\alpha_2\beta_2,\alpha_3\beta_3}_{3^{+-}(j=2)} &=& \mathcal{S}[ \bar q_a \sigma^{\alpha_1\beta_1} \lambda_n^{ab} q_b~d^{npq}~g_s^2 G_p^{\alpha_2\beta_2} G_q^{\alpha_3\beta_3}] \, ,
\label{def:23pm}
\\ 
J^{\alpha\beta}_{1^{--}(j=0)} &=& \bar q_a \sigma^{\alpha\beta} \lambda_n^{ab} q_b~d^{npq}~g_s^2 G_p^{\mu\nu} \tilde G_{q,\mu\nu} \, ,
\label{def:01mm}
\\
J_{0^{--}(j=1)} &=& \bar q_a \sigma^{\mu\nu} \lambda_n^{ab} q_b~d^{npq}~g_s^2 G_{p,\nu\rho} \tilde G_{q,\mu}^\rho \, ,
\label{def:10mm}
\\
J^{\alpha\beta}_{1^{--}(j=1)} &=& \bar q_a \sigma_{\alpha_1\beta_1} \lambda_n^{ab} q_b~d^{npq}~g_s^2 \left( G_{p,\alpha_2\mu} \tilde G_{q,\beta_2}^{\mu} - \{ \alpha_2 \leftrightarrow \beta_2 \} \right) \times g^{\beta_1 \beta_2} (g^{\alpha\alpha_1}g^{\beta\alpha_2} - g^{\beta\alpha_1}g^{\alpha\alpha_2} ) \, ,
\label{def:11mm}
\\
J^{\alpha_1\beta_1,\alpha_2\beta_2}_{2^{--}(j=1)} &=& \mathcal{S}[ \bar q_a \sigma^{\alpha_1\beta_1} \lambda_n^{ab} q_b~d^{npq}~g_s^2 \left( G_p^{\alpha_2\mu} \tilde G_{q,\mu}^{\beta_2} - \{ \alpha_2 \leftrightarrow \beta_2 \} \right) ] \, ,
\label{def:12mm}
\\
J^{\alpha\beta}_{1^{--}(j=2)} &=& \bar q_a \sigma^{\mu_2\nu_2} \lambda_n^{ab} q_b~d^{npq}~\mathcal{S}^\prime[ g_s^2 G_{p,\mu_1\nu_1} \tilde G_{q,\mu_2\nu_2}] \times g^{\alpha\mu_1}g^{\beta\nu_1} \, ,
\label{def:21mm}
\\
J^{\alpha_1\beta_1,\alpha_2\beta_2}_{2^{--}(j=2)} &=&  \mathcal{S}[ \bar q_a \sigma_{\mu_3\nu_3} \lambda_n^{ab} q_b~d^{npq}~\mathcal{S}^\prime[ g_s^2 G_{p,\mu_1\nu_1} \tilde G_{q,\mu_2\nu_2}] \times g^{\alpha_1\mu_1}g^{\beta_1\nu_1}g^{\nu_2\nu_3}(g^{\alpha_2\mu_2}g^{\beta_2\mu_3} - g^{\beta_2\mu_2}g^{\alpha_2\mu_3})] \, ,
\label{def:22mm}
\\
J^{\alpha_1\beta_1,\alpha_2\beta_2,\alpha_3\beta_3}_{3^{--}(j=2)} &=& \mathcal{S}[ \bar q_a \sigma^{\alpha_1\beta_1} \lambda_n^{ab} q_b~d^{npq}~g_s^2 G_p^{\alpha_2\beta_2} \tilde G_q^{\alpha_3\beta_3}] \, .
\label{def:23mm}
\end{eqnarray}
\end{widetext}
In the above expressions, $\mathcal{S}$ represents symmetrization and subtracting trace terms in the two sets $\{\alpha_1\cdots\alpha_J\}$ and $\{\beta_1\cdots\beta_J\}$ simultaneously, and $\mathcal{S}^\prime$ represents symmetrization and subtracting trace terms in the two sets $\{\mu_1\mu_2\}$ and $\{\nu_1\nu_2\}$ simultaneously. However, we do not know the explicit expressions of these two rearrangements, so we are not able to calculate the decay constants, and moreover, we are not able to calculate the eight currents $J^{\cdots}_{1/2^{\pm\pm}(j=2)}$ in the present study.

The color-octet quark-antiquark field $\bar q_a \lambda_n^{ab} \sigma_{\mu\nu} q_b$ has the spin-parity quantum number $J^P = 1^\pm$. Due to its $S$-wave component of $J^P = 1^-$, the above double-gluon hybrid currents may couple to the lowest-lying double-gluon hybrid states. As shown in Fig.~\ref{fig:category}, the four currents $J^{\alpha\beta}_{1^{\pm\pm}(j=0)}$ contain the double-gluon fields of $j=0$:
\begin{equation}
G_p^{\mu\nu} G_{q,\mu\nu} \, , \, G_p^{\mu\nu} \tilde G_{q,\mu\nu} \, ;
\end{equation}
the twelve currents $J^{\cdots}_{0/1/2^{\pm\pm}(j=1)}$ contain the double-gluon fields of $j=1$:
\begin{equation}
G_{p,\mu\rho} G_{q,\nu}^{\rho} - \{ \mu \leftrightarrow \nu \} \, , \, G_{p,\mu\rho} \tilde G_{q,\nu}^{\rho} - \{ \mu \leftrightarrow \nu \} \, ;
\end{equation}
the twelve currents $J^{\cdots}_{1/2/3^{\pm\pm}(j=2)}$ contain the double-gluon fields of $j=2$:
\begin{equation}
\mathcal{S}^\prime[ G_{p,\mu_1\nu_1} G_{q,\mu_2\nu_2}] \, , \, \mathcal{S}^\prime[ G_{p,\mu_1\nu_1} \tilde G_{q,\mu_2\nu_2}] \, .
\end{equation}
Especially, the double-gluon hybrid currents $J^{\alpha\beta}_{1^{-+}(j=0/1/2)}$ and $J^{\alpha_1\beta_1,\alpha_2\beta_2,\alpha_3\beta_3}_{3^{-+}(j=2)}$ with the exotic quantum numbers $J^{PC} = 1^{-+}$ and $3^{-+}$ are of particular interest, since these quantum numbers can not be reached by the conventional $\bar q q$ mesons.

Each of the four currents, $J_{0^{\pm\pm}(j=1)}$, can only couple to either the positive- or negative-parity state:
\begin{equation}
\langle 0 | J_{0^{\pm\pm}(j=1)} | X ; 0^{\pm\pm}(j=1) \rangle = f_{0^{\pm\pm}(j=1)} \, .
\end{equation}
The other twenty-four currents with non-zero spins $J \neq 0$, $J^{\cdots}_{1/2/3^{\pm\pm}(j=0/1/2)}$, all have $2 \times J$ Lorentz indices with certain symmetries. Each of them can couple to both the positive- and negative-parity states. As already discussed in Refs.~\cite{Chen:2021smz,Su:2022fqr}, we can write the couplings of these currents to the double-gluon hybrid states $| X ; J^{PC}(j)\rangle$ as:
\begin{eqnarray}
&& \langle 0 | J^{\alpha\beta}_{1^{\pm\pm}(j=0/1/2)} | X ; 1^{\pm\pm}(j=0/1/2) \rangle
\label{eq:1oh}
\\ \nonumber &=& i f_{1^{\pm\pm}(j=0/1/2)} ~ \epsilon_\mu p_\nu ~ \epsilon^{\alpha\beta\mu\nu} \, ,
\\ && \langle 0 | J^{\alpha_1\beta_1,\alpha_2\beta_2}_{2^{\pm\pm}(j=1/2)} | X ; 2^{\pm\pm}(j=1/2) \rangle
\\ \nonumber &=& i f_{2^{\pm\pm}(j=1/2)} ~ \epsilon_{\mu_1 \mu_2} p_{\nu_1} p_{\nu_2} ~ \mathcal{S}[ \epsilon^{\alpha_1 \beta_1 \mu_1 \nu_1} \epsilon^{\alpha_2 \beta_2 \mu_2 \nu_2} ] \, ,
\label{eq:3oh}
\\ && \langle 0 | J^{\alpha_1\beta_1,\alpha_2\beta_2,\alpha_3\beta_3}_{3^{\pm\pm}(j=2)} | X ; 3^{\pm\pm}(j=2) \rangle
\\ \nonumber &=& i f_{3^{\pm\pm}(j=2)} ~ \epsilon_{\mu_1 \mu_2 \mu_3} p_{\nu_1} p_{\nu_2} p_{\nu_3}
\\ \nonumber && \times \mathcal{S}[ \epsilon^{\alpha_1 \beta_1 \mu_1 \nu_1} \epsilon^{\alpha_2 \beta_2 \mu_2 \nu_2} \epsilon^{\alpha_3 \beta_3 \mu_3 \nu_3} ] \, .
\end{eqnarray}
Here $| X ; 1^{\pm\pm}(j=0/1/2) \rangle $, $| X ; 2^{\pm\pm}(j=1/2) \rangle $, and $| X ; 3^{\pm\pm}(j=2) \rangle $ are the double-gluon hybrid states that have the same parities as the components $J^{ij}_{1^{\pm\pm}(j=0/1/2)}$, $J^{i_1j_1,i_2j_2}_{2^{\pm\pm}(j=1/2)}$, and $J^{i_1j_1,i_2j_2,i_3j_3}_{3^{\pm\pm}(j=2)}$ ($i,i_1,i_2,i_3,j,j_1,j_2,j_3=1,2,3$), respectively.

Before performing QCD sum rule analyses in the next section, we further examine the twenty-eight currents $J^{\cdots}_{0/1/2/3^{\pm\pm}(j=0/1/2)}$. We find that some internal symmetries between the two gluon fields make the eleven currents $J^{\cdots}_{0^{\pm-}(j=1),1^{\pm+}(j=0),1^{\pm-}(j=1),2^{\pm-}(j=1),1/2/3^{++}(j=2)}$ vanish:
\begin{itemize}

\item The current $J^{\alpha\beta}_{1^{++}(j=0)}$ contains two totally antisymmetric gluon fields (their color indices are antisymmetric and their Lorentz indices are symmetric), so it is zero due to the Bose-Einstein statistics:
\begin{eqnarray}
\nonumber J^{\alpha\beta}_{1^{++}(j=0)} &=& \cdots \times f^{npq} ~ G_p^{\mu\nu} G_{q,\mu\nu}
\\ \nonumber &=& \cdots \times f^{nqp} ~ G_q^{\mu\nu} G_{p,\mu\nu}
\\ \nonumber &=& - \cdots \times f^{npq} ~ G_p^{\mu\nu} G_{q,\mu\nu}
\\ &=& - J^{\alpha\beta}_{1^{++}(j=0)} \, .
\end{eqnarray}

\item The current $J^{\alpha\beta}_{1^{-+}(j=0)}$ is zero because
\begin{eqnarray}
\nonumber J^{\alpha\beta}_{1^{-+}(j=0)} &=& \cdots \times f^{npq} ~ G_p^{\mu\nu} \tilde G_{q,\mu\nu}
\\ \nonumber &=& \cdots \times f^{nqp} ~ G_q^{\mu\nu} \tilde G_{p,\mu\nu}
\\ \nonumber &=& - \cdots \times f^{npq} ~ \tilde G_p^{\mu\nu} G_{q,\mu\nu}
\\ \nonumber &=& - \cdots \times f^{npq} ~ G_p^{\mu\nu} \tilde G_{q,\mu\nu}
\\ &=& - J^{\alpha\beta}_{1^{-+}(j=0)} \, .
\end{eqnarray}

\item The three currents $J^{\cdots}_{0/1/2^{+-}(j=1)}$ all contain two totally antisymmetric gluon fields (their color indices are symmetric and their Lorentz indices are antisymmetric), so they are zero due to the Bose-Einstein statistics:
\begin{eqnarray}
\nonumber J^{\cdots}_{0/1/2^{+-}(j=1)} &=& \cdots \times d^{npq}~\left( G_p^{\alpha\mu} G_{q,\mu}^\beta - G_p^{\beta\mu} G_{q,\mu}^\alpha \right)
\\ \nonumber &=& \cdots \times d^{nqp}~\left( G_q^{\alpha\mu} G_{p,\mu}^\beta - G_q^{\beta\mu} G_{p,\mu}^\alpha \right)
\\ \nonumber &=& - \cdots \times d^{npq}~\left( G_p^{\alpha\mu} G_{q,\mu}^\beta - G_p^{\beta\mu} G_{q,\mu}^\alpha \right)
\\ &=& - J^{\cdots}_{0/1/2^{+-}(j=1)} \, .
\label{eq:identity1}
\end{eqnarray}

\item The three currents $J^{\cdots}_{0/1/2^{--}(j=1)}$ are zero because
\begin{eqnarray}
\nonumber && d^{npq}~\left( G_p^{0\mu} \tilde G_q^{1\mu} - G_p^{1\mu} \tilde G_q^{0\mu} \right)
\\ \nonumber &=& d^{npq}~\left( G_p^{02} \tilde G_q^{12} + G_p^{03} \tilde G_q^{13} - G_p^{12} \tilde G_q^{02} - G_p^{13} \tilde G_q^{03} \right)
\\ \nonumber &=& {d^{npq} \over2}~\left( G_p^{02} G_q^{03} - G_p^{03} G_q^{02} + G_p^{12} G_q^{13} - G_p^{13} G_q^{12} \right)
\\ &=& 0 \, .
\label{eq:identity2}
\end{eqnarray}

\item The three currents $J^{\cdots}_{1/2/3^{++}(j=2)}$ all contain two totally antisymmetric gluon fields (their color indices are antisymmetric and their Lorentz indices are symmetric), so they are zero due to the Bose-Einstein statistics, {\it e.g.},
\begin{eqnarray}
\nonumber J^{\alpha_1\beta_1,\alpha_2\beta_2,\alpha_3\beta_3}_{3^{++}(j=2)} &=& \cdots \times f^{npq} ~ \mathcal{S}[ G_p^{\alpha_1\beta_1} G_q^{\alpha_2\beta_2} ]
\\ \nonumber &=& \cdots \times f^{nqp} ~ \mathcal{S}[ G_q^{\alpha_1\beta_1} G_p^{\alpha_2\beta_2} ]
\\ \nonumber &=& - \cdots \times f^{npq} ~ \mathcal{S}[ G_p^{\alpha_2\beta_2} G_q^{\alpha_1\beta_1} ]
\\ \nonumber &=& - \cdots \times f^{npq} ~ \mathcal{S}[ G_p^{\alpha_1\beta_1} G_q^{\alpha_2\beta_2} ]
\\ &=& - J^{\alpha_1\beta_1,\alpha_2\beta_2,\alpha_3\beta_3}_{3^{++}(j=2)} \, .
\end{eqnarray}

\end{itemize}
According to Yang and Landau's theorem~\cite{Landau:1948kw,Yang:1950rg}, the states with $J=1$ do not decay into two photons, and one can not construct an interpolating current with $J=1$ using two electromagnetic field strength tensors. Similarly, we have the color-singlet glueball operators, such as $G^n_{\mu\nu} G_n^{\mu\nu}$ and $G^n_{\mu\nu} {\tilde G_n}^{\mu\nu}$, with $n$ the color index; however, we can not construct the glueball interpolating current with $J=1$ which is composed of two gluon field strength tensors with the same color indices. We note that the color indices of the gluon fields in each of the six currents in Eqs.~(\ref{eq:identity1}) and (\ref{eq:identity2}) are fully symmetric due to the color factor $d^{npq}$. Hence, all these currents with the spin of the double-gluon fields $j=1$ vanish, which is the manifestation of the generalized Yang and Landau's theorem in QCD.

%
\section{QCD sum rule analyses}
\label{sec:sumrule}
%

In this section we apply the QCD sum rule method to study the eleven currents $J^{\cdots}_{0^{\pm+}(j=1)/1^{\pm-}(j=0)/1^{\pm+}(j=1)/2^{\pm+}(j=1)/3^{\pm-}(j=2)/3^{-+}(j=2)}$. This method has been widely applied to study various exotic hadrons~\cite{Shifman:1978bx,Reinders:1984sr,Chen:2020aos,Chen:2020uif,Chen:2021erj}. We use the current $J^{\alpha \beta}_{1^{-+}(j=1)}$ given in Eq.~(\ref{def:11mp}) as an example, which couples to the state $|X;1^{-+}(j=1)\rangle$ through Eq.~(\ref{eq:1oh}). We shall calculate its two-point correlation function
\begin{eqnarray}
\nonumber && \Pi^{\alpha \beta;\alpha^\prime \beta^\prime}(q^2)
\\ \nonumber &=& i \int d^4x e^{iqx} \langle 0 | {\bf T}[J^{\alpha \beta}_{1^{-+}(j=1)}(x) J^{\alpha^\prime \beta^\prime, \dagger}_{1^{-+}(j=1)}(0)] | 0 \rangle
\\ &=& (g^{\alpha\alpha^\prime}g^{\beta\beta^\prime}-g^{\alpha\beta^\prime}g^{\beta\alpha^\prime})~\Pi(q^2)
\, ,
\label{eq:correlation}
\end{eqnarray}
at both the hadron and quark-gluon levels.

At the hadron level, we use the dispersion relation to describe Eq.~(\ref{eq:correlation}) as
%
\begin{equation}
\Pi(q^2) = \int^\infty_{s_<}\frac{\rho(s)}{s-q^2-i\varepsilon}ds \, ,
\label{eq:hadron}
\end{equation}
%
where $s_<$ is the physical threshold, and $\rho(s) \equiv {\rm Im}\Pi(s)/\pi$ is the spectral density. We parameterize $\rho(s)$ as one pole dominance for the possibly-existing ground state $X \equiv | X; 1^{-+}(j=1) \rangle$ and a continuum contribution
%
\begin{eqnarray}
&&\nonumber \rho_{\rm phen}(s)
\\ \nonumber&\equiv& \sum_n\delta(s-M^2_n) \langle 0| J^{\alpha \beta}_{1^{-+}(j=1)} | n\rangle \langle n| J^{\alpha^\prime \beta^\prime, \dagger}_{1^{-+}(j=1)} |0 \rangle
\\ &=& f^2_X \delta(s-M^2_X) + \rm{continuum} \, .
\label{eq:rho}
\end{eqnarray}
%

At the quark-gluon level, we use the method of operator product expansion (OPE) to calculate Eq.~(\ref{eq:correlation}), and extract the spectral density $\rho_{\rm OPE}(s)$. Then the Borel transformation is carried out at both the hadron and quark-gluon levels, and the continuum contribution is approximated as the integral above the threshold value $s_0$:
%
\begin{equation}
\Pi(s_0, M_B^2) \equiv f^2_X e^{-M_X^2/M_B^2} = \int^{s_0}_{s_<} e^{-s/M_B^2}\rho_{\rm OPE}(s)ds \, .
\label{eq:fin}
\end{equation}
%
Finally, the mass of $|X;1^{-+}(j=1) \rangle$ is deduced to be
%
\begin{equation}
M^2_X(s_0, M_B) = \frac{\int^{s_0}_{s_<} e^{-s/M_B^2}s\rho_{\rm OPE}(s)ds}{\int^{s_0}_{s_<} e^{-s/M_B^2}\rho_{\rm OPE}(s)ds} \, .
\label{eq:LSR}
\end{equation}

In this paper we have calculated $\rho_{\rm OPE}(s)$ up to the dimension eight ($D=8$) condensates, including the perturbative term, the quark condensates $\langle \bar q q \rangle$ and $\langle \bar s s \rangle$, the two-gluon condensate $\langle g_s^2 G^2 \rangle$, the three-gluon condensate $\langle g_s^3 G^3 \rangle$, the quark-gluon mixed condensates $\langle g_s \bar q \sigma G q \rangle$ and $\langle g_s \bar s \sigma G s \rangle$, and their combinations. Besides, we have taken into account all the terms proportional to $\alpha_s^2 \times g_s^0 = g_s^4$ and $\alpha_s^2 \times g_s^1 = g_s^5$, while we have partly taken into account the terms proportional to $\alpha_s^2 \times g_s^2 = g_s^6$.

The spectral density extracted from the current $J^{\alpha \beta}_{1^{-+}(j=1)}$ with the quark-gluon content $q \bar q g g $ ($q=u/d$) is
\begin{eqnarray}
\rho_{1^{-+}(j=1)}^{\bar{q}qgg}(s) &=& \frac{\alpha_s^2 s^5}{5040 \pi^4}
\label{rho:q1pm}
+\frac{17\alpha_s^2 \langle g_s^2 GG \rangle s^3}{4608 \pi^4}
\\ \nonumber &&-
\frac{\alpha_s \langle g_s^3 G^3 \rangle s^2}{48\pi^3}  -4 \alpha_s^2 \langle \bar s s \rangle \langle g_s \bar s \sigma G s \rangle s \, .
\end{eqnarray}
Note that the double-gluon hybrid states within the same isospin multiplet have the same extracted hadron mass when using the QCD sum rule method, because we do not differentiate the up and down quarks. The corresponding one with the quark-gluon content $\bar s s gg$ is
\begin{eqnarray}
&& \rho_{1^{-+}(j=1)}^{\bar{s}sgg}(s)
\label{rho:s1pm}
\\ \nonumber &=& \frac{\alpha_s^2 s^5}{5040 \pi^4}- \frac{ \alpha_s^2 m_s^2 s^4}{90 \pi^4}
\\ \nonumber &&+
\left(  \frac{17\alpha_s^2 \langle g_s^2 GG \rangle }{4608 \pi^4}
+ \frac{2 \alpha_s^2 m_s \langle \bar s s \rangle }{15 \pi^2}
+ \frac{\alpha_s^2 m_s^4 }{10 \pi^4}\right) s^3
\\ \nonumber &&+ \left( - \frac{ \alpha_s \langle g_s^3 G^3 \rangle}{48\pi^3}-\frac{4 \alpha_s^2 m_s^3  \langle \bar s s \rangle}{3\pi^2}
\right.
\\ \nonumber &&+
\left.
\frac{ \alpha_s^2 m_s  \langle g_s \bar s \sigma G s \rangle}{4\pi^2} - \frac{ \alpha_s^2 m_s^2  \langle g_s^2 GG \rangle}{64\pi^4} \right) s^2
\\ \nonumber &&+ \left( -4\alpha_s^2 \langle \bar s s \rangle \langle g_s \bar s \sigma G s \rangle +\frac{ 5\alpha_s m_s^2 \langle g_s^3 G^3 \rangle}{12\pi^3}
\right.
\\ \nonumber &&+
\left.
\frac{16 \alpha_s^2 m_s^2 \langle \bar s s \rangle^2}{9}
-\frac{23 \alpha_s^2 m_s \langle g_s^2 GG \rangle \langle \bar s s \rangle}{48\pi^2}
\right.
\\ \nonumber &&+
\left.
\frac{5 \alpha_s^2 m_s^4 \langle g_s^2 GG \rangle }{64\pi^4} \right) s \, ,
\end{eqnarray}
Other spectral densities can be found in Appendix~\ref{app:sumrule}.

\section{Numerical Analyses}
\label{sec:numerical}

In this section we perform numerical analyses to study the spectral densities listed in Eqs.~(\ref{rho:q1pm}) and (\ref{rho:s1pm}) as well as those listed in Appendix~\ref{app:sumrule}. Here are various QCD parameters at the renormalization scale 2 GeV and the QCD scale $\Lambda_{\rm QCD} = 300$~MeV~\cite{pdg,Ovchinnikov:1988gk,Yang:1993bp,Ellis:1996xc,Ioffe:2002be,Jamin:2002ev,Gimenez:2005nt,Narison:2011xe,Narison:2018dcr}:
\begin{eqnarray}
\nonumber \alpha_s(Q^2) &=& {4\pi \over 11 \ln(Q^2/\Lambda_{\rm QCD}^2)} \, ,
\\ \nonumber m_s &=& 93 ^{+11}_{-5} \mbox{ MeV} \, ,
\\ \nonumber \langle\bar qq \rangle &=& -(0.240 \pm 0.010)^3 \mbox{ GeV}^3 \, ,
\\ \langle\bar ss \rangle &=& (0.8\pm 0.1)\times \langle\bar qq \rangle \, ,
\label{eq:condensate}
\\  \nonumber \langle g_s\bar q\sigma G q\rangle &=& (0.8 \pm 0.2)\times\langle\bar qq\rangle \mbox{ GeV}^2 \, ,
\\ \nonumber \langle g_s\bar s\sigma G s\rangle &=&  (0.8 \pm 0.2)\times\langle\bar ss\rangle \, ,
\\ \nonumber \langle \alpha_s GG\rangle &=& (6.35 \pm 0.35) \times 10^{-2} \mbox{ GeV}^4 \, ,
\\ \nonumber \langle g_s^3G^3\rangle &=& (8.2 \pm 1.0) \times \langle \alpha_s GG\rangle  \mbox{ GeV}^2 \, .
\end{eqnarray}

We use the spectral density $\rho^{\bar q q gg}_{1^{-+}(j=1)}(s)$ given in Eq.~(\ref{rho:q1pm}) as an example, which is extracted from the current $J^{\alpha \beta}_{1^{-+}(j=1)}$ with the quark-gluon content $\bar q q g g$ ($q=u/d$). This current couples to the state $| \bar q q gg; 1^{-+}(j=1) \rangle$, whose mass depends on two free parameters: the threshold value $s_0$ and the Borel mass $M_B$. We shall find their proper working regions through three steps.

\begin{figure}[hbtp]
\begin{center}
\includegraphics[width=0.45\textwidth]{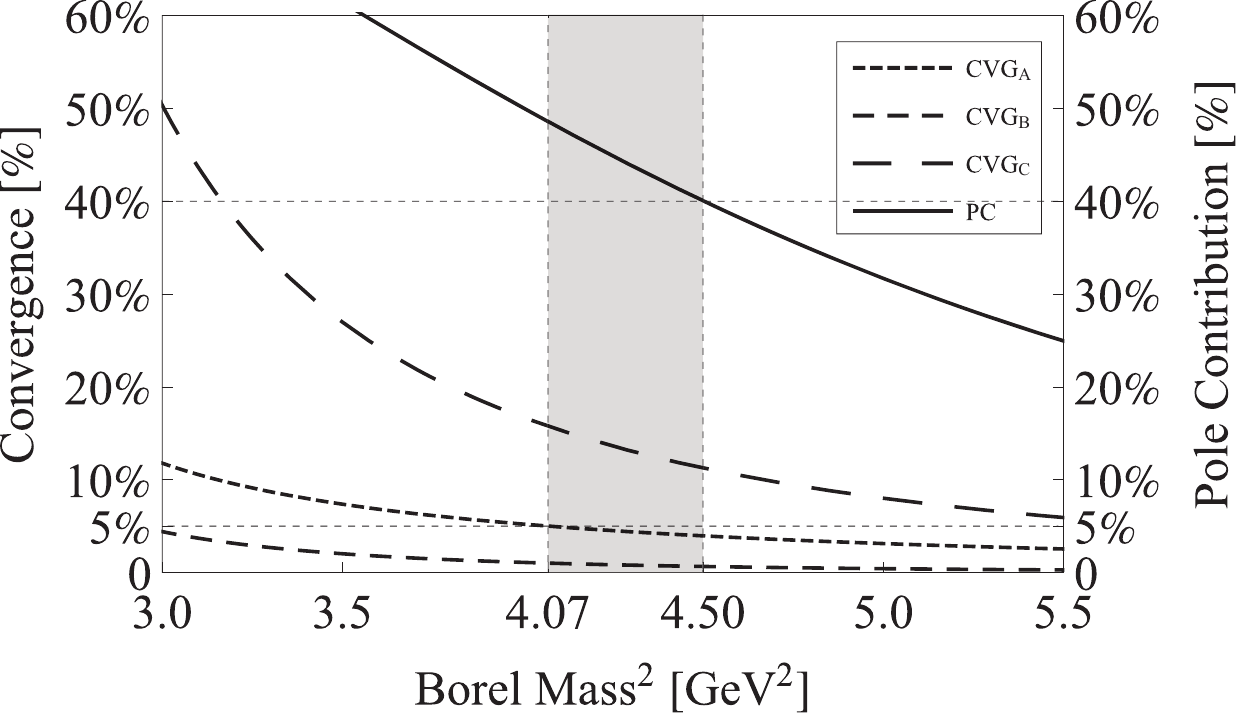}
\caption{CVG$_{A/B/C}$ and PC as functions of the Borel mass $M_B$, when setting $s_0 = 24.0$~GeV$^2$. These curves are obtained using the spectral density $\rho^{\bar q q gg}_{1^{-+}(j=1)}(s)$, which is extracted from the current $J^{\alpha \beta}_{1^{-+}(j=1)}$ with the quark-gluon content $\bar q q g g$ ($q=u/d$).}
\label{fig:cvgpole}
\end{center}
\end{figure}

Firstly, as the cornerstone of QCD sum rule analyses, the convergence of OPE is satisfied by requiring: a) the $\alpha_s^2 \times g_s^2 = g_s^6$ terms are less than 5\%, b) the $D=8$ terms are less than 10\%, and c) the $D=6$ terms are less than 20\%:
\begin{eqnarray}
\mbox{CVG}_A &\equiv& \left|\frac{ \Pi^{g_s^{n=6}}(\infty, M_B^2) }{ \Pi(\infty, M_B^2) }\right| \leq 5\% \, ,\label{eq:convergence1}
\\
\mbox{CVG}_B &\equiv& \left|\frac{ \Pi^{{\rm D=8}}(\infty, M_B^2) }{ \Pi(\infty, M_B^2) }\right| \leq 10\% \, ,\label{eq:convergence2}
\\
\mbox{CVG}_C &\equiv& \left|\frac{ \Pi^{D=6}(\infty, M_B^2) }{ \Pi(\infty, M_B^2) }\right| \leq 20\% \, .\label{eq:convergence3}
\end{eqnarray}
As shown in Fig.~\ref{fig:cvgpole} through three dashed curves, we obtain that the lower bound of the Borel mass is $M_B^2 \geq 4.07$~GeV$^2$.

Secondly, the one-pole-dominance assumption is satisfied by requiring the pole contribution (PC) to be larger than $40\%$:
\begin{equation}
\mbox{PC} \equiv \left|\frac{ \Pi(s_0, M_B^2) }{ \Pi(\infty, M_B^2) }\right| \geq 40\% \, .
\label{eq:pole}
\end{equation}
As shown in Fig.~\ref{fig:cvgpole} through the solid curve, we obtain that the upper bound of the Borel mass is $M_B^2 \leq 4.50$~GeV$^2$, when setting $s_0 = 24.0$~GeV$^2$.

\begin{figure*}[]
\begin{center}
\subfigure[(a)]{\includegraphics[width=0.4\textwidth]{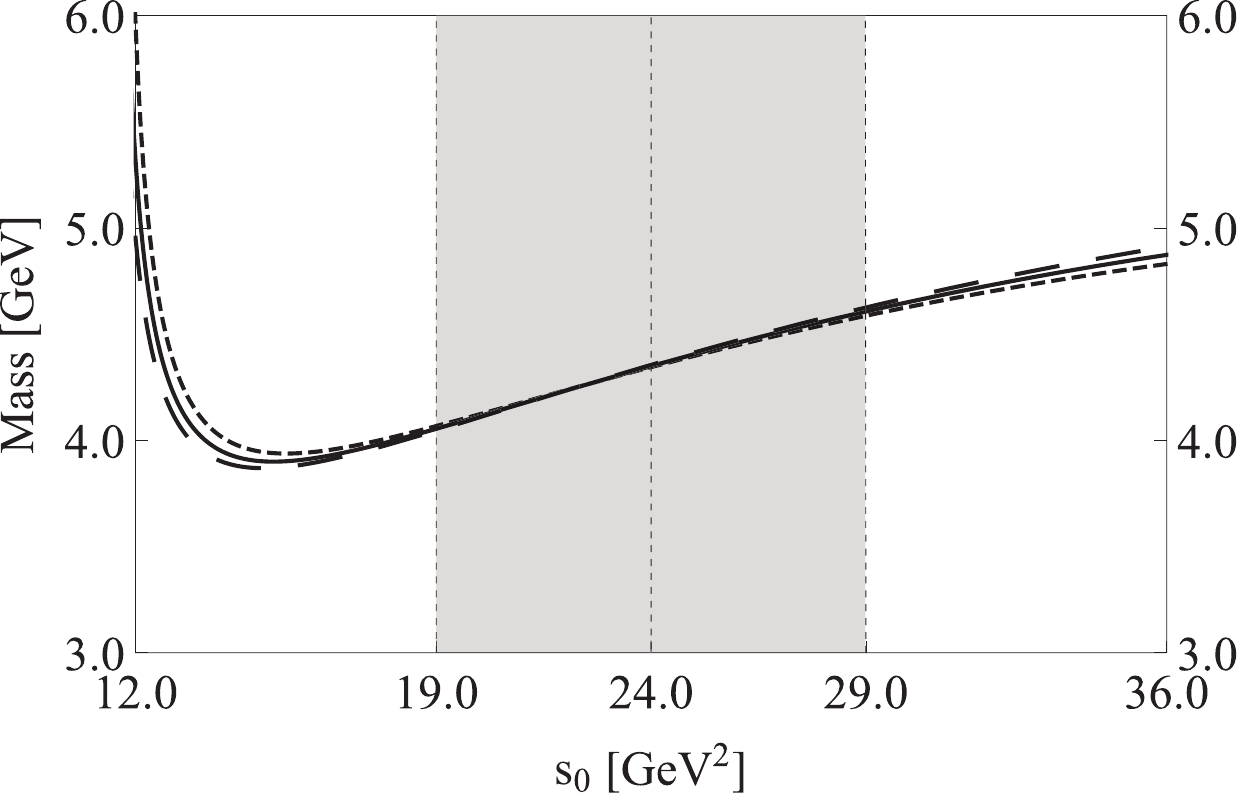}}
~~~~~~~~~~
\subfigure[(b)]{\includegraphics[width=0.4\textwidth]{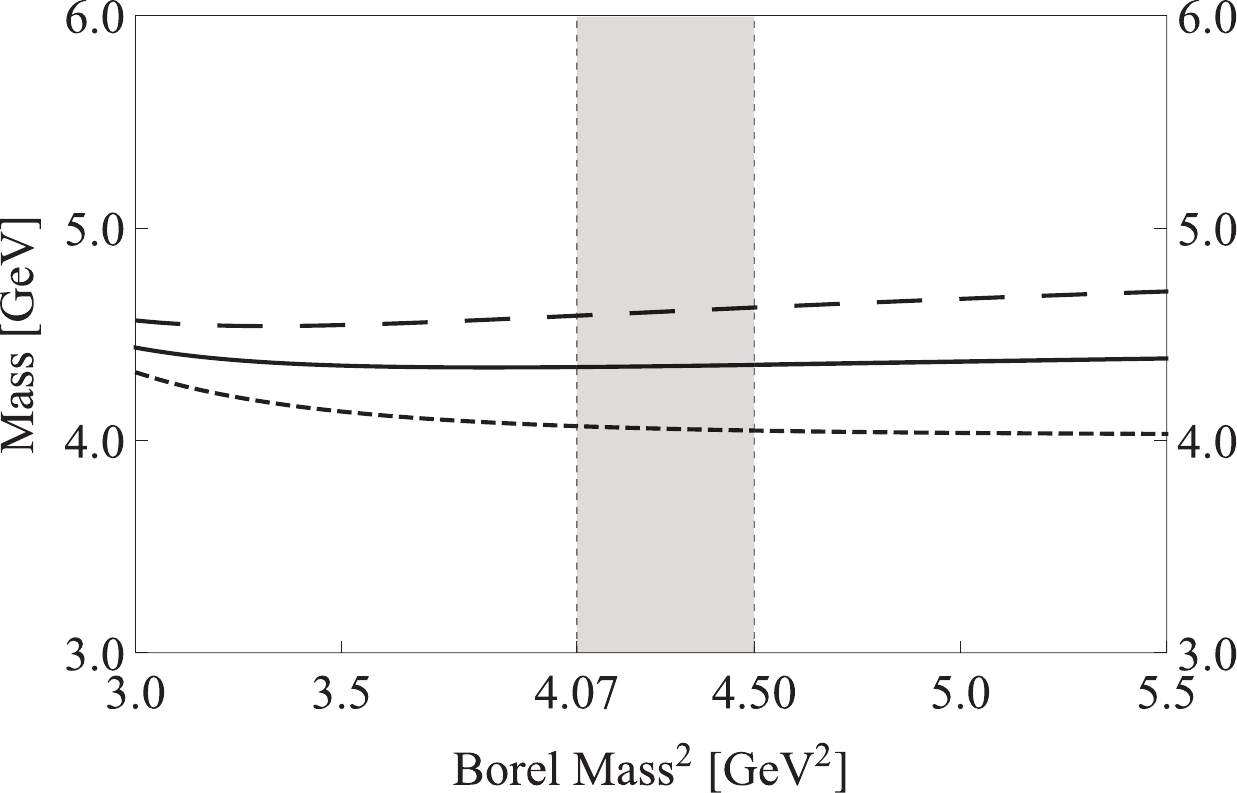}}
\caption{Mass of the double-gluon hybrid state $|\bar q q gg;1^{-+}(j=1)\rangle$ with respect to (a) the threshold value $s_0$ and (b) the Borel mass $M_B$. In the subfigure (a) the short-dashed/middle-dashed/long-dashed curves are obtained by setting $M_B^2 = 4.07/4.29/4.50$~GeV$^2$, respectively. In the subfigure (b) the short-dashed/middle-dashed/long-dashed curves are obtained by setting $s_0 = 19.0/24.0/29.0$~GeV$^2$, respectively. These curves are obtained using the spectral density $\rho^{\bar q q gg}_{1^{-+}(j=1)}(s)$, which is extracted from the current $J^{\alpha \beta}_{1^{-+}(j=1)}$ with the quark-gluon content $\bar q q g g$ ($q=u/d$).}
\label{fig:mass}
\end{center}
\end{figure*}

Therefore, we obtain the Borel window to be $4.07$~GeV$^2 \leq M_B^2 \leq 4.50$~GeV$^2$ when setting $s_0 = 24.0$~GeV$^2$. We change $s_0$ to repeat the same procedures, and find that there are non-vanishing Borel windows when setting $s_0 > s_{0}^{min} = 22.1$~GeV$^2$. We choose $s_0$ to be slightly larger, and determine its working region to be $19.0$~GeV$^2 \leq s_0 \leq 29.0$~GeV$^2$.

Thirdly, we investigate the mass dependence on $s_0$ and $M_B^2$. As shown in Fig.~\ref{fig:mass}(a), we find a mass minimum around $s_0 \sim 15$~GeV$^2$, and the mass dependence on $s_0$ is moderate and acceptable inside the region $19.0$~GeV$^2 \leq s_0 \leq 29.0$~GeV$^2$. As shown in Fig.~\ref{fig:mass}(b), the mass dependence on $M_B$ is weak inside the Borel window $4.07$~GeV$^2 < M_B^2 < 4.50$~GeV$^2$.

From the above three steps, we determine our working regions to be $19.0$~GeV$^2 \leq s_0 \leq 29.0$~GeV$^2$ and $4.07$~GeV$^2 \leq M_B^2 \leq 4.50$~GeV$^2$, where we calculate the mass of the state $| \bar q q gg; 1^{-+}(j=1) \rangle$ to be
\begin{equation}
M_{|\bar q q gg;1^{-+}(j=1)\rangle} = 4.35^{+0.26}_{-0.30}{\rm~GeV} \, .
\end{equation}
Its central value is obtained using $s_0 = 24.0$~GeV$^2$ and $M_B^2 = 4.29$~GeV$^2$, and its uncertainty comes from the threshold value $s_0$, the Borel mass $M_B$, and various QCD parameters listed in Eqs.~(\ref{eq:condensate}).

Similarly, we perform numerical analyses to study the other spectral densities. The obtained results are summarized in Table~\ref{tab:results}. Especially, the masses of the double-gluon hybrid states $|\bar s s gg;1^{-+}(j=1)\rangle$, $|\bar q q gg;3^{-+}(j=2)\rangle$, and $|\bar s s gg;3^{-+}(j=2)\rangle$ are calculated to be
\begin{eqnarray}
M_{|\bar s s gg;1^{-+}(j=1)\rangle} &=& 4.49^{+0.25}_{-0.30}{\rm~GeV} \, ,
\\
M_{|\bar q q gg;3^{-+}(j=2)\rangle} &=& 3.02^{+0.24}_{-0.31}{\rm~GeV} \, ,
\\
M_{|\bar s s gg;3^{-+}(j=2)\rangle} &=& 3.16^{+0.22}_{-0.28}{\rm~GeV} \, .
\end{eqnarray}

%
\section{Summary and Discussions}
\label{sec:summary}

\begin{table*}[]
\begin{center}
\renewcommand{\arraystretch}{1.6}
\caption{QCD sum rule results of the double-gluon hybrid states $|\bar q q gg; J^{PC}(j)\rangle$ and $|\bar s s gg; J^{PC}(j)\rangle$, extracted from the eleven currents $J^{\cdots}_{0^{\pm+}(j=1)/1^{\pm-}(j=0)/1^{\pm+}(j=1)/2^{\pm+}(j=1)/3^{\pm-}(j=2)/3^{-+}(j=2)}$ with the quark-gluon contents $\bar q q g g$ ($q=u/d$) and $\bar s s g g$. Their partner states with the quark-gluon content $\bar q s g g$, denoted as $|\bar q s gg; J^{P(C)}(j)\rangle$, are also studied for completeness. We use $J$ to denote the total spin of the double-gluon hybrid current, and $j$ to denote the spin of the double-gluon field.}
\begin{tabular}{c|c|c|c|c|c|c}
\hline\hline
~~~\multirow{2}{*}{State [$J^{PC}$]}~~~ & ~~~~\multirow{2}{*}{Current}~~~~ & ~\multirow{2}{*}{~$s_0^{min}~[{\rm GeV}^2]$~}~ & \multicolumn{2}{c|}{Working Regions} & ~~\multirow{2}{*}{Pole~[\%]}~~ & ~\multirow{2}{*}{~Mass~[GeV]~}~
\\ \cline{4-5}
&  &  & ~~$M_B^2~[{\rm GeV}^2]$~~ & ~~$s_0~[{\rm GeV}^2]$~~ &&
\\ \hline\hline
$|\bar q q gg; 0^{++}(j=1)\rangle$ & $J_{0^{++}(j=1)}$                                         &  27.0   &  $5.13$--$5.76$   &  $30\pm6.0$  &  $40$--$50$  &  $4.76^{+0.30}_{-0.37}$
\\
$|\bar q q gg; 0^{-+}(j=1)\rangle$ & $J_{0^{-+}(j=1)}$                                         &  27.3   &  $5.16$--$5.74$   &  $30\pm6.0$  &  $40$--$49$  &  $4.79^{+0.30}_{-0.36}$
\\
$|\bar q q gg; 1^{+-}(j=0)\rangle$ & $J^{\alpha\beta}_{1^{+-}(j=0)}$                           &  26.0   &  $5.37$--$5.93$   &  $29\pm6.0$  &  $40$--$50$  &  $4.50^{+0.34}_{-0.44}$
\\
$|\bar q q gg; 1^{--}(j=0)\rangle$ & $J^{\alpha\beta}_{1^{--}(j=0)}$                           &  33.3   &  $5.91$--$6.82$   &  $37\pm7.0$  &  $40$--$52$  &  $5.45^{+0.28}_{-0.30}$
\\
$|\bar q q gg; 1^{++}(j=1)\rangle$ & $J^{\alpha\beta}_{1^{++}(j=1)}$                           &  21.7   &  $4.03$--$4.54$   &  $24\pm4.0$  &  $40$--$50$  &  $4.31^{+0.22}_{-0.26}$
\\
$|\bar q q gg; 1^{-+}(j=1)\rangle$ & $J^{\alpha\beta}_{1^{-+}(j=1)}$                           &  22.1   &  $4.07$--$4.50$   &  $24\pm5.0$  &  $40$--$49$  &  $4.35^{+0.26}_{-0.30}$
\\
$|\bar q q gg; 2^{++}(j=1)\rangle$ & $J^{\alpha_1\beta_1,\alpha_2\beta_2}_{2^{++}(j=1)}$       &  17.9    &  $4.00$--$4.34$   &  $20\pm4.0$   &  $40$--$48$  &  $3.60^{+0.30}_{-0.38}$
\\
$|\bar q q gg; 2^{-+}(j=1)\rangle$ & $J^{\alpha_1\beta_1,\alpha_2\beta_2}_{2^{-+}(j=1)}$       &  30.0   &  $5.24$--$6.02$   &  $33\pm7.0$  &  $40$--$51$  &  $5.21^{+0.27}_{-0.26}$
\\
$|\bar q q gg; 3^{+-}(j=2)\rangle$ & $J^{\alpha_1\beta_1,\alpha_2\beta_2,\alpha_3\beta_3}_{3^{+-}(j=2)}$       &  13.4   &  $3.57$--$3.89$   &  $15\pm3.0$  &  $40$--$48$  &  $3.12^{+0.24}_{-0.30}$
\\
$|\bar q q gg; 3^{--}(j=2)\rangle$ & $J^{\alpha_1\beta_1,\alpha_2\beta_2,\alpha_3\beta_3}_{3^{--}(j=2)}$       &  17.9    &  $4.65$--$5.03$   &  $20\pm4.0$   &  $40$--$48$  &  $3.59^{+0.28}_{-0.35}$
\\
$|\bar q q gg; 3^{-+}(j=2)\rangle$ & $J^{\alpha_1\beta_1,\alpha_2\beta_2,\alpha_3\beta_3}_{3^{-+}(j=2)}$       &  11.6    &  $2.68$--$2.95$   &  $13\pm3.0$  &  $40$--$49$  &  $3.02^{+0.24}_{-0.31}$
\\[1mm] \hline\hline
$|\bar s s gg; 0^{++}(j=1)\rangle$ & $J_{0^{++}(j=1)}$                                         &  27.5   &  $5.17$--$6.13$   &  $32\pm6.0$  &  $40$--$55$  &  $4.91^{+0.29}_{-0.35}$
\\
$|\bar s s gg; 0^{-+}(j=1)\rangle$ & $J_{0^{-+}(j=1)}$                                         &  28.0   &  $5.23$--$6.10$   &  $32\pm6.0$  &  $40$--$54$  &  $4.93^{+0.28}_{-0.35}$
\\
$|\bar s s gg; 1^{+-}(j=0)\rangle$ & $J^{\alpha\beta}_{1^{+-}(j=0)}$                           &  26.7   &  $5.43$--$6.21$   &  $31\pm6.0$  &  $40$--$53$  &  $4.67^{+0.33}_{-0.42}$
\\
$|\bar s s gg; 1^{--}(j=0)\rangle$ & $J^{\alpha\beta}_{1^{--}(j=0)}$                           &  33.8   &  $5.88$--$7.15$   &  $39\pm8.0$  &  $40$--$56$  &  $5.59^{+0.30}_{-0.33}$
\\
$|\bar s s gg; 1^{++}(j=1)\rangle$ & $J^{\alpha\beta}_{1^{++}(j=1)}$                           &  22.2   &  $4.05$--$4.91$   &  $26\pm5.0$  &  $40$--$57$  &  $4.47^{+0.25}_{-0.31}$
\\
$|\bar s s gg; 1^{-+}(j=1)\rangle$ & $J^{\alpha\beta}_{1^{-+}(j=1)}$                           &  22.8   &  $4.16$--$4.88$   &  $26\pm5.0$  &  $40$--$54$  &  $4.49^{+0.25}_{-0.30}$
\\
$|\bar s s gg; 2^{++}(j=1)\rangle$ & $J^{\alpha_1\beta_1,\alpha_2\beta_2}_{2^{++}(j=1)}$       &  18.2    &  $4.00$--$4.44$   &  $21\pm4.0$   &  $40$--$51$  &  $3.70^{+0.30}_{-0.37}$
\\
$|\bar s s gg; 2^{-+}(j=1)\rangle$ & $J^{\alpha_1\beta_1,\alpha_2\beta_2}_{2^{-+}(j=1)}$       &  30.5   &  $5.21$--$6.36$   &  $35\pm7.0$  &  $40$--$56$  &  $5.35^{+0.27}_{-0.26}$
\\
$|\bar s s gg; 3^{+-}(j=2)\rangle$ & $J^{\alpha_1\beta_1,\alpha_2\beta_2,\alpha_3\beta_3}_{3^{+-}(j=2)}$       &  13.0   &  $3.45$--$4.04$   &  $16\pm3.0$  &  $40$--$55$  &  $3.23^{+0.23}_{-0.28}$
\\
$|\bar s s gg; 3^{--}(j=2)\rangle$ & $J^{\alpha_1\beta_1,\alpha_2\beta_2,\alpha_3\beta_3}_{3^{--}(j=2)}$       &  17.7    &  $4.57$--$5.17$   &  $21\pm4.0$   &  $40$--$52$  &  $3.68^{+0.28}_{-0.33}$
\\
$|\bar s s gg; 3^{-+}(j=2)\rangle$ & $J^{\alpha_1\beta_1,\alpha_2\beta_2,\alpha_3\beta_3}_{3^{-+}(j=2)}$       &  11.6    &  $2.57$--$3.06$   &  $14\pm3.0$  &  $40$--$56$  &  $3.16^{+0.22}_{-0.28}$
\\[1mm] \hline\hline
$|\bar q s gg; 0^{+(+)}(j=1)\rangle$ & $J_{0^{+(+)}(j=1)}$                                         &  27.3   &  $5.17$--$5.95$   &  $31\pm6.0$  &  $40$--$53$  &  $4.84^{+0.30}_{-0.36}$
\\
$|\bar q s gg; 0^{-(+)}(j=1)\rangle$ & $J_{0^{-(+)}(j=1)}$                                         &  27.6   &  $5.19$--$5.92$   &  $31\pm6.0$  &  $40$--$52$  &  $4.86^{+0.29}_{-0.35}$
\\
$|\bar q s gg; 1^{+(-)}(j=0)\rangle$ & $J^{\alpha\beta}_{1^{+(-)}(j=0)}$                           &  26.3   &  $5.40$--$6.08$   &  $30\pm6.0$  &  $40$--$51$  &  $4.59^{+0.34}_{-0.43}$
\\
$|\bar q s gg; 1^{-(-)}(j=0)\rangle$ & $J^{\alpha\beta}_{1^{-(-)}(j=0)}$                           &  33.5   &  $5.88$--$6.87$   &  $38\pm8.0$  &  $40$--$54$  &  $5.52^{+0.30}_{-0.33}$
\\
$|\bar q s gg; 1^{+(+)}(j=1)\rangle$ & $J^{\alpha\beta}_{1^{+(+)}(j=1)}$                           &  21.9   &  $4.04$--$4.72$   &  $25\pm5.0$  &  $40$--$54$  &  $4.38^{+0.26}_{-0.31}$
\\
$|\bar q s gg; 1^{-(+)}(j=1)\rangle$ & $J^{\alpha\beta}_{1^{-(+)}(j=1)}$                           &  22.5   &  $4.12$--$4.69$   &  $25\pm5.0$  &  $40$--$51$  &  $4.42^{+0.25}_{-0.30}$
\\
$|\bar q s gg; 2^{+(+)}(j=1)\rangle$ & $J^{\alpha_1\beta_1,\alpha_2\beta_2}_{2^{+(+)}(j=1)}$       &  18.0    &  $4.00$--$4.39$   &  $20.5\pm4.0$   &  $40$--$50$  &  $3.65^{+0.30}_{-0.37}$
\\
$|\bar q s gg; 2^{-(+)}(j=1)\rangle$ & $J^{\alpha_1\beta_1,\alpha_2\beta_2}_{2^{-(+)}(j=1)}$       &  30.2   &  $5.21$--$6.19$   &  $34\pm7.0$  &  $40$--$54$  &  $5.28^{+0.27}_{-0.26}$
\\
$|\bar q s gg; 3^{+(-)}(j=2)\rangle$ & $J^{\alpha_1\beta_1,\alpha_2\beta_2,\alpha_3\beta_3}_{3^{+(-)}(j=2)}$       &  13.0   &  $3.48$--$3.96$   &  $15.5\pm3.0$  &  $40$--$52$  &  $3.17^{+0.24}_{-0.29}$
\\
$|\bar q s gg; 3^{-(-)}(j=2)\rangle$ & $J^{\alpha_1\beta_1,\alpha_2\beta_2,\alpha_3\beta_3}_{3^{-(-)}(j=2)}$       &  17.7    &  $4.59$--$5.10$   &  $20.5\pm4.0$   &  $40$--$50$  &  $3.64^{+0.28}_{-0.34}$
\\
$|\bar q s gg; 3^{-(+)}(j=2)\rangle$ & $J^{\alpha_1\beta_1,\alpha_2\beta_2,\alpha_3\beta_3}_{3^{-(+)}(j=2)}$       &  11.3    &  $2.56$--$3.01$   &  $13.5\pm3.0$  &  $40$--$55$  &  $3.08^{+0.23}_{-0.29}$
\\[1mm] \hline\hline
\end{tabular}
\label{tab:results}
\end{center}
\end{table*}

In this paper we study the double-gluon hybrid states, whose corresponding interpolating currents are constructed using the color-octet quark-antiquark field $ \bar q_a \sigma_{\mu \nu} \lambda_n^{ab} q_b$ together with the color-octet double-gluon fields $d^{npq} G_p^{\alpha\beta} G_q^{\gamma\delta}$ and $f^{npq} G_p^{\alpha\beta} G_q^{\gamma\delta}$. We systematically construct twenty-eight currents, eleven of which are found to be zero due to some internal symmetries between the two gluon fields, partly as the manifestation of the generalized Yang and Landau's theorem in QCD~\cite{Landau:1948kw,Yang:1950rg}. We apply the QCD sum rule method to study the rest of the currents, but some of them are not so easy due to the complicated Lorentz indices.

Altogether we calculate eleven double-gluon hybrid currents in the present study, and their quark-gluon contents are assumed to be either $\bar q q gg$ ($q=u/d$) or $\bar s s gg$. The obtained QCD sum rule results are summarized in Table~\ref{tab:results}. For completeness, we denote their partner states with the quark-gluon content $\bar q s g g$ as $|\bar q s gg; J^{P(C)}(j)\rangle$, whose results are also calculated and summarized in Table~\ref{tab:results}. These currents have the quantum numbers $J^{PC}(j)=0^{\pm+}(j=1)/1^{\pm-}(j=0)/1^{\pm+}(j=1)/2^{\pm+}(j=1)/3^{\pm-}(j=2)/3^{-+}(j=2)$, with $j$ the spin of the double-gluon field. Especially, they can reach the exotic quantum numbers $J^{PC}=1^{-+}$ and $3^{-+}$ that the traditional $\bar q q$ mesons can not reach. Note that these two exotic quantum numbers have not been investigated in our previous studies~\cite{Chen:2021smz,Su:2022fqr}.

As shown in Table~\ref{tab:results}, the masses of the double-gluon hybrid states with $J^{PC}=1^{-+}$ and $3^{-+}$ are calculated to be:
\begin{eqnarray}
\nonumber M_{|\bar q q gg;1^{-+}(j=1)\rangle} &=& 4.35^{+0.26}_{-0.30}{\rm~GeV} \, ,
\\ \nonumber M_{|\bar s s gg;1^{-+}(j=1)\rangle} &=& 4.49^{+0.25}_{-0.30}{\rm~GeV} \, ,
\\ \nonumber M_{|\bar q q gg;3^{-+}(j=2)\rangle} &=& 3.02^{+0.24}_{-0.31}{\rm~GeV} \, ,
\\ \nonumber M_{|\bar s s gg;3^{-+}(j=2)\rangle} &=& 3.16^{+0.22}_{-0.28}{\rm~GeV} \, .
\end{eqnarray}
Because we do not differentiate the up and down quarks in the present study, the states within the same isospin multiplet have the same mass, {\it i.e.},
\begin{eqnarray}
\nonumber M_{|\bar q q gg;0^+ 1^{-+}(j=1)\rangle}    &=& 4.35^{+0.26}_{-0.30}{\rm~GeV} \, ,
\\ \nonumber M_{|\bar q q gg;1^- 1^{-+}(j=1)\rangle} &=& 4.35^{+0.26}_{-0.30}{\rm~GeV} \, ,
\\ \nonumber M_{|\bar s s gg;0^+ 1^{-+}(j=1)\rangle} &=& 4.49^{+0.25}_{-0.30}{\rm~GeV} \, ,
\\ \nonumber M_{|\bar q q gg;0^+3^{-+}(j=2)\rangle}  &=& 3.02^{+0.24}_{-0.31}{\rm~GeV} \, ,
\\ \nonumber M_{|\bar q q gg;1^-3^{-+}(j=2)\rangle}  &=& 3.02^{+0.24}_{-0.31}{\rm~GeV} \, ,
\\ \nonumber M_{|\bar s s gg;0^+3^{-+}(j=2)\rangle}  &=& 3.16^{+0.22}_{-0.28}{\rm~GeV} \, .
\end{eqnarray}
Especially, the masses of the $J^{PC} = 3^{-+}$ states are calculated to be slightly larger than 3.0~GeV, which values are accessible in the BESIII, Belle-II, GlueX, LHCb, and PANDA experiments, etc.

To end this paper, we briefly discuss possible decay patterns of these double-gluon hybrid states. Generally speaking, they can decay after exciting two $\bar qq/\bar ss$ pairs from two gluons and then reorganizing into two or three color-singlet light mesons. We list in Table~\ref{tab:decay} some possible decay patterns for the double-gluon hybrid states $|\bar q q gg;0^+3^{-+}(j=2)\rangle$, $|\bar q q gg;1^-3^{-+}(j=2)\rangle$, and $|\bar s s gg;0^+3^{-+}(j=2)\rangle$. Accordingly, we propose to search for $|\bar q q gg;0^+3^{-+}(j=2)\rangle$ in the $\pi a_1(1260)$ channel, to search for $|\bar q q gg;1^-3^{-+}(j=2)\rangle$ in the $\rho \omega$ channel, and to search for $|\bar s s gg;0^+3^{-+}(j=2)\rangle$ in the $\phi \phi$ channel, etc.

\begin{table*}[]
\begin{center}
\renewcommand{\arraystretch}{1.5}
\caption{Possible two- and three-meson decay patterns of the $J^{PC}=3^{-+}$ double-gluon hybrid states with the quark-gluon contents $\bar q q gg$ ($q=u/d$) and $\bar s s gg$. The following notations are used here: $a_0 = a_0(980)$, $f_0 = f_0(980)$, $a_1 = a_1(1260)$, $b_1 = b_1(1235)$, $h_1 = h_1(1170)$, $f_1 = f_1(1285)$, $h_1^\prime = h_1(1415)$, $f_1^\prime = f_1(1420)$, $a_2 = a_2(1320)$, $K_0^* = K_0^*(700)$, $K_1 = K_1(1270)/K_1(1400)$, and $K_2^* = K_2^*(1430)$.}
\begin{tabular}{ c | c | c | c }
\hline\hline
Two-Meson               & ~~~~~~~~~~~$|\bar q q gg;0^+3^{-+}(j=2)\rangle$~~~~~~~~~~~                                 & ~~~~~~~~$|\bar q q gg;1^-3^{-+}(j=2)\rangle$~~~~~~~~               & ~~~~~~~~$|\bar s s gg;0^+3^{-+}(j=2)\rangle$~~~~~~~~
\\ \hline\hline
{$S$-wave} & \multicolumn{3}{c}{$K_2^*\bar K^* $}
\\ \hline
\multirow{2}{*}{$P$-wave} &   --    & $\rho \omega$                  & $\phi \phi$
\\ \cline{2-4}
                        & \multicolumn{3}{c}{$K_2^* \bar K_1 , K_2^* \bar K_0^* $}
\\ \hline
\multirow{2}{*}{$D$-wave} & $\pi a_1, \pi a_2, \rho b_1$                               & $\pi f_1, \pi f_2, \rho h_1$                   & $\phi h_1^\prime, \eta^{(\prime)} f_1^\prime$
\\ \cline{2-4}
                        & \multicolumn{3}{c}{$K_1 \bar K, K_2^* \bar K, K_1 \bar K^*, K_2^* \bar K^*, K^* \bar K_0^*$}
\\ \hline\hline
~Three-Meson~           & ~~~~~~~~~~~$|\bar q q gg;0^+3^{-+}(j=2)\rangle$~~~~~~~~~~~                                 & ~~~~~~~~$|\bar q q gg;1^-3^{-+}(j=2)\rangle$~~~~~~~~               & ~~~~~~~~$|\bar s s gg;0^+3^{-+}(j=2)\rangle$~~~~~~~~
\\ \hline\hline
{$S$-wave}                &    \multicolumn{3}{c}{$\rho K^*\bar K^* $}
\\ \hline
\multirow{3}{*}{$P$-wave} & $ \pi \pi f_2, \rho \rho f_0, \eta^{(\prime)} \eta^{(\prime)} f_2$            & $ \pi \pi a_2, \rho \rho a_0, \eta^{(\prime)} \eta^{(\prime)} a_2$         & \multirow{2}{*}{$ \phi \eta^{(\prime)} h_1^\prime$}
\\ & $\pi \eta^{(\prime)} a_2, \pi \rho h_1 ,\pi \omega b_1$            & $\pi \rho b_1, \pi \eta^{(\prime)} f_2 ,\pi \omega h_1$         &
\\ \cline{2-4}
                        & \multicolumn{3}{c}{$\rho K_1 \bar K, \omega K_1 \bar K, \pi K_1 \bar K^*, \eta^{(\prime)} K_1 \bar K^*, \phi K_1 \bar K, \phi K_2^* \bar K$}
\\ \hline
\multirow{2}{*}{$D$-wave} & $\rho \rho \eta^{(\prime)} , \omega \omega \eta^{(\prime)}, \pi \rho \omega$                               & $\rho \rho \pi , \omega \omega \pi $                    & $\phi \phi \eta^{(\prime)}$
\\ \cline{2-4}
                        & \multicolumn{3}{c}{$\rho K \bar K, \pi K^* \bar K^*, \pi K^* \bar K, \eta^{(\prime)} K^* \bar K, \rho K^* \bar K,\omega K^* \bar K,\phi K^* \bar K^*, \eta^{(\prime)} K^* \bar K^*, \phi K^* \bar K$}
\\ \hline\hline
\end{tabular}
\label{tab:decay}
\end{center}
\end{table*}

\section*{Acknowledgments}
%

This project is supported by
the National Natural Science Foundation of China under Grant No. 11722540, No.~11975033, No. 12075019, and No.~12070131001,
the National Key R$\&$D Program of China under Contracts No. 2020YFA0406400,
the Jiangsu Provincial Double-Innovation Program under Grant No.~JSSCRC2021488,
and
the Fundamental Research Funds for the Central Universities.

\appendix
\section{Spectral densities}
\label{app:sumrule}

In this appendix we list the spectral densities extracted from the ten double-gluon hybrid currents $J^{\cdots}_{0^{\pm+}(j=1)/1^{\pm-}(j=0)/1^{++}(j=1)/2^{\pm+}(j=1)/3^{\pm-}(j=2)/3^{-+}(j=2)}$ with the quark-gluon contents $\bar q q gg$ ($q=u/d$) and $\bar s s gg$, as follows:
\begin{widetext}
\begin{eqnarray}
\rho_{0^{++}(j=1)}^{\bar{q}qgg}(s) &=& \frac{\alpha_s^2 s^5}{2400 \pi^4} + \frac{121\alpha_s^2 \langle g_s^2 GG \rangle s^3}{9216 \pi^4} - \frac{ \alpha_s \langle g_s^3 G^3 \rangle s^2}{16\pi^3}  + 12\alpha_s^2 \langle \bar q q \rangle \langle g_s \bar q \sigma G q \rangle s \, ,
\\
\rho_{0^{++}(j=1)}^{\bar{s}sgg}(s) &=& \frac{\alpha_s^2 s^5}{2400 \pi^4}- \frac{\alpha_s^2 m_s^2 s^4}{40 \pi^4}  + \left(  \frac{121\alpha_s^2 \langle g_s^2 GG \rangle }{9216 \pi^4} + \frac{\alpha_s^2 m_s \langle \bar s s \rangle }{3 \pi^2} + \frac{\alpha_s^2 m_s^4 }{4 \pi^4}\right) s^3
\\ \nonumber &&+ \left( - \frac{ \alpha_s \langle g_s^3 G^3 \rangle}{16\pi^3}-\frac{4 \alpha_s^2 m_s^3  \langle \bar s s \rangle}{\pi^2} -\frac{3 \alpha_s^2 m_s  \langle g_s \bar s \sigma G s \rangle}{2\pi^2}-\frac{69 \alpha_s^2 m_s^2  \langle g_s^2 GG \rangle}{256\pi^4} \right) s^2
\\ \nonumber && + \left( 12\alpha_s^2 \langle \bar s s \rangle \langle g_s \bar s \sigma G s \rangle  + 8 \alpha_s^2 m_s^2 \langle \bar s s \rangle^2 +\frac{3 \alpha_s m_s^2 \langle g_s^3 G^3 \rangle}{8\pi^3} +\frac{57 \alpha_s^2 m_s \langle g_s^2 GG \rangle \langle \bar s s \rangle}{32\pi^2}+\frac{45 \alpha_s^2 m_s^4 \langle g_s^2 GG \rangle }{128\pi^4} \right) s \, ,
\\
\rho_{0^{-+}(j=1)}^{\bar{q}qgg}(s) &=& \frac{\alpha_s^2 s^5}{2400 \pi^4} + \frac{121\alpha_s^2 \langle g_s^2 GG \rangle s^3}{9216 \pi^4}  - \frac{ \alpha_s \langle g_s^3 G^3 \rangle s^2}{16\pi^3}  - 12\alpha_s^2 \langle \bar q q \rangle \langle g_s \bar q \sigma G q \rangle s \, ,
\\
\rho_{0^{-+}(j=1)}^{\bar{s}sgg}(s) &=& \frac{\alpha_s^2 s^5}{2400 \pi^4}- \frac{\alpha_s^2 m_s^2 s^4}{40 \pi^4}  + \left(  \frac{121\alpha_s^2 \langle g_s^2 GG \rangle }{9216 \pi^4} + \frac{\alpha_s^2 m_s \langle \bar s s \rangle }{3 \pi^2} + \frac{\alpha_s^2 m_s^4 }{ 4\pi^4}\right) s^3
\\ \nonumber &&+ \left( - \frac{ \alpha_s \langle g_s^3 G^3 \rangle}{16\pi^3}-\frac{4 \alpha_s^2 m_s^3  \langle \bar s s \rangle}{\pi^2} +\frac{3 \alpha_s^2 m_s  \langle g_s \bar s \sigma G s \rangle}{2\pi^2}-\frac{39 \alpha_s^2 m_s^2  \langle g_s^2 GG \rangle}{256\pi^4} \right) s^2
\\ \nonumber && + \left( - 12\alpha_s^2 \langle \bar s s \rangle \langle g_s \bar s \sigma G s \rangle  + 8 \alpha_s^2 m_s^2 \langle \bar s s \rangle^2 +\frac{9 \alpha_s m_s^2 \langle g_s^3 G^3 \rangle}{8\pi^3} -\frac{3 \alpha_s^2 m_s \langle g_s^2 GG \rangle \langle \bar s s \rangle}{32\pi^2}+\frac{45 \alpha_s^2 m_s^4 \langle g_s^2 GG \rangle }{128\pi^4} \right) s \, ,
\\
\rho_{1^{++}(j=1)}^{\bar{q}qgg}(s) &=& \frac{\alpha_s^2 s^5}{5040 \pi^4} + \frac{17\alpha_s^2 \langle g_s^2 GG \rangle s^3}{4608 \pi^4}  - \frac{ \alpha_s \langle g_s^3 G^3 \rangle s^2}{48\pi^3}  + 4\alpha_s^2 \langle \bar q q \rangle \langle g_s \bar q \sigma G q \rangle s \, ,
\\
\rho_{1^{++}(j=1)}^{\bar{s}sgg}(s) &=& \frac{\alpha_s^2 s^5}{5040 \pi^4}- \frac{\alpha_s^2 m_s^2 s^4}{90 \pi^4}  + \left(  \frac{17\alpha_s^2 \langle g_s^2 GG \rangle }{4608 \pi^4} + \frac{2 \alpha_s^2 m_s \langle \bar s s \rangle }{15 \pi^2} + \frac{\alpha_s^2 m_s^4 }{10 \pi^4}\right) s^3
\\ \nonumber &&+ \left( - \frac{ \alpha_s \langle g_s^3 G^3 \rangle}{48\pi^3}-\frac{4 \alpha_s^2 m_s^3  \langle \bar s s \rangle}{3\pi^2} -\frac{ \alpha_s^2 m_s  \langle g_s \bar s \sigma G s \rangle}{2\pi^2} - \frac{3 \alpha_s^2 m_s^2  \langle g_s^2 GG \rangle}{32\pi^4} \right) s^2
\\ \nonumber &&+ \left( 4\alpha_s^2 \langle \bar s s \rangle \langle g_s \bar s \sigma G s \rangle +\frac{ \alpha_s m_s^2 \langle g_s^3 G^3 \rangle}{12\pi^3}+\frac{16 \alpha_s^2 m_s^2 \langle \bar s s \rangle^2}{9}+\frac{37 \alpha_s^2 m_s \langle g_s^2 GG \rangle \langle \bar s s \rangle}{48\pi^2}+\frac{5 \alpha_s^2 m_s^4 \langle g_s^2 GG \rangle }{64\pi^4} \right) s \, ,
\\
\rho_{1^{+-}(j=0)}^{\bar{q}qgg}(s) &=& \frac{\alpha_s^2 s^5}{6048 \pi^4} + \frac{23\alpha_s^2 \langle g_s^2 GG \rangle s^3}{3456 \pi^4}+  \left(  \frac{ 5 \alpha_s \langle g_s^3 G^3 \rangle}{288\pi^3}-\frac{ 80 \alpha_s^2 \langle \bar q q \rangle^2}{27}\right) s^2
\\ \nonumber &&+ \left( \frac{80\alpha_s^2 \langle \bar q q \rangle \langle g_s \bar q \sigma G q \rangle}{9}+\frac{5\langle g_s^2 GG \rangle^2}{864\pi^2}\right)\, ,
\\
\rho_{1^{+-}(j=0)}^{\bar{s}sgg}(s) &=& \frac{\alpha_s^2 s^5}{6048 \pi^4}- \frac{7 \alpha_s^2 m_s^2 s^4}{432 \pi^4}  + \left(  \frac{23\alpha_s^2 \langle g_s^2 GG \rangle }{3456 \pi^4} + \frac{13 \alpha_s^2 m_s \langle \bar s s \rangle }{27 \pi^2} + \frac{\alpha_s^2 m_s^4 }{12 \pi^4}\right) s^3
\\ \nonumber &&+ \left(  \frac{ 5 \alpha_s \langle g_s^3 G^3 \rangle}{288\pi^3}-\frac{ 80 \alpha_s^2 \langle \bar s s \rangle^2}{27}-\frac{10 \alpha_s^2 m_s^3  \langle \bar s s \rangle}{9\pi^2} -\frac{ 10\alpha_s^2 m_s  \langle g_s \bar s \sigma G s \rangle}{9\pi^2} - \frac{ 25\alpha_s^2 m_s^2  \langle g_s^2 GG \rangle}{96\pi^4} \right) s^2
\\ \nonumber &&+ \left( \frac{80\alpha_s^2 \langle \bar s s \rangle \langle g_s \bar s \sigma G s \rangle}{9}+\frac{5\langle g_s^2 GG \rangle^2}{864\pi^2} -\frac{ 25\alpha_s m_s^2 \langle g_s^3 G^3 \rangle}{72\pi^3}+\frac{40 \alpha_s^2 m_s^2 \langle \bar s s \rangle^2}{27}\right.
\\ \nonumber &&\left.
+\frac{175 \alpha_s^2 m_s \langle g_s^2 GG \rangle \langle \bar s s \rangle}{72\pi^2}+\frac{25 \alpha_s^2 m_s^4 \langle g_s^2 GG \rangle }{96\pi^4} \right) s \, ,
\\
\rho_{1^{--}(j=0)}^{\bar{q}qgg}(s) &=& \frac{\alpha_s^2 s^5}{6048 \pi^4} + \frac{\alpha_s^2 \langle g_s^2 GG \rangle s^3}{6912 \pi^4}  + \left( - \frac{ 5\alpha_s \langle g_s^3 G^3 \rangle}{96\pi^3}- \frac{ 80\alpha_s^2  \langle \bar q q \rangle^2}{27}\right)s^2
\\ \nonumber &&+ \left( \frac{80\alpha_s^2 \langle \bar q q \rangle \langle g_s \bar q \sigma G q \rangle}{9} -\frac{5 \langle g_s^2 GG \rangle^2 }{864\pi^2}\right)s\, ,
\\
\rho_{1^{--}(j=0)}^{\bar{s}sgg}(s) &=& \frac{\alpha_s^2 s^5}{6048 \pi^4}- \frac{7 \alpha_s^2 m_s^2 s^4}{432 \pi^4}  + \left(  \frac{\alpha_s^2 \langle g_s^2 GG \rangle }{6912 \pi^4} + \frac{52 \alpha_s^2 m_s \langle \bar s s \rangle }{108 \pi^2} + \frac{\alpha_s^2 m_s^4 }{12 \pi^4}\right) s^3
\\ \nonumber &&+ \left( - \frac{ 5\alpha_s \langle g_s^3 G^3 \rangle}{96\pi^3}- \frac{ 80\alpha_s^2  \langle \bar s s \rangle^2}{27}-\frac{10 \alpha_s^2 m_s^3  \langle \bar s s \rangle}{9\pi^2} -\frac{ 10\alpha_s^2 m_s  \langle g_s \bar s \sigma G s \rangle}{9\pi^2}  \right) s^2
\\ \nonumber &&+ \left( \frac{80\alpha_s^2 \langle \bar s s \rangle \langle g_s \bar s \sigma G s \rangle}{9} -\frac{5 \langle g_s^2 GG \rangle^2 }{864\pi^2}+\frac{ 25\alpha_s m_s^2 \langle g_s^3 G^3 \rangle}{24\pi^3}+\frac{40 \alpha_s^2 m_s^2 \langle \bar s s \rangle^2}{27} \right) s \, ,
\\
\rho_{2^{++}(j=1)}^{\bar{q}qgg}(s) &=& \frac{\alpha_s^2 s^5}{28800 \pi^4} + \frac{65\alpha_s^2 \langle g_s^2 GG \rangle s^3 }{82944 \pi^4}  + \left( \frac{ \alpha_s \langle g_s^3 G^3 \rangle}{320\pi^3}- \frac{ 4\alpha_s^2  \langle \bar q q \rangle^2}{9}\right)s^2
\\ \nonumber &&+ \left( \frac{8\alpha_s^2 \langle \bar q q \rangle \langle g_s \bar q \sigma G q \rangle}{9} +\frac{ \langle g_s^2 GG \rangle^2 }{576\pi^2}\right)s\, ,
\\
\rho_{2^{++}(j=1)}^{\bar{s}sgg}(s) &=& \frac{\alpha_s^2 s^5}{28800 \pi^4}- \frac{\alpha_s^2 m_s^2 s^4}{315 \pi^4}  + \left(  \frac{65\alpha_s^2 \langle g_s^2 GG \rangle }{82944 \pi^4} + \frac{23 \alpha_s^2 m_s \langle \bar s s \rangle }{270 \pi^2} + \frac{\alpha_s^2 m_s^4 }{72 \pi^4}\right) s^3
\\ \nonumber &&+ \left(  \frac{ \alpha_s \langle g_s^3 G^3 \rangle}{320\pi^3}- \frac{ 4\alpha_s^2  \langle \bar s s \rangle^2}{9}-\frac{2 \alpha_s^2 m_s^3  \langle \bar s s \rangle}{15\pi^2} -\frac{ \alpha_s^2 m_s  \langle g_s \bar s \sigma G s \rangle}{6\pi^2} -\frac{ 29\alpha_s^2 m_s^2\langle g_s^2 GG \rangle}{1024\pi^4} \right) s^2
\\ \nonumber &&+ \left( \frac{8\alpha_s^2 \langle \bar s s \rangle \langle g_s \bar s \sigma G s \rangle}{9} +\frac{ \langle g_s^2 GG \rangle^2 }{576\pi^2}-\frac{ \alpha_s m_s^2 \langle g_s^3 G^3 \rangle}{12\pi^3}+\frac{15 \alpha_s^2 m_s \langle \bar s s \rangle \langle g_s^2 GG \rangle}{64\pi^2} +\frac{5 \alpha_s^2 m_s^4 \langle g_s^2 GG \rangle}{256\pi^4}\right) s \, ,
\\
\rho_{2^{-+}(j=1)}^{\bar{q}qgg}(s) &=& \frac{\alpha_s^2 s^5}{28800 \pi^4} + \frac{\alpha_s^2 \langle g_s^2 GG \rangle s^3}{5184 \pi^4}  + \left(  \frac{ 7\alpha_s \langle g_s^3 G^3 \rangle}{960\pi^3}- \frac{ 4\alpha_s^2  \langle \bar q q \rangle^2}{9}\right)s^2
\\ \nonumber &&+ \left( \frac{8\alpha_s^2 \langle \bar q q \rangle \langle g_s \bar s \sigma G q \rangle}{9} -\frac{ \langle g_s^2 GG \rangle^2 }{576\pi^2}\right)s\, ,
\\
\rho_{2^{-+}(j=1)}^{\bar{s}sgg}(s) &=& \frac{\alpha_s^2 s^5}{28800 \pi^4}- \frac{ \alpha_s^2 m_s^2 s^4}{315 \pi^4}  + \left(  \frac{\alpha_s^2 \langle g_s^2 GG \rangle }{5184 \pi^4} + \frac{46 \alpha_s^2 m_s \langle \bar s s \rangle }{540 \pi^2} + \frac{\alpha_s^2 m_s^4 }{72 \pi^4}\right) s^3
\\ \nonumber &&+ \left(  \frac{ 7\alpha_s \langle g_s^3 G^3 \rangle}{960\pi^3}- \frac{ 4\alpha_s^2  \langle \bar s s \rangle^2}{9}-\frac{2 \alpha_s^2 m_s^3  \langle \bar s s \rangle}{15\pi^2} -\frac{ 2\alpha_s^2 m_s  \langle g_s \bar s \sigma G s \rangle}{12\pi^2}+\frac{11\alpha_s^2 m_s^2 \langle g_s^2 GG \rangle}{1024\pi^4}  \right) s^2
\\ \nonumber &&+ \left( \frac{8\alpha_s^2 \langle \bar s s \rangle \langle g_s \bar s \sigma G s \rangle}{9} -\frac{ \langle g_s^2 GG \rangle^2 }{576\pi^2}+\frac{ \alpha_s m_s^2 \langle g_s^3 G^3 \rangle}{8\pi^3}-\frac{25 \alpha_s^2 m_s \langle \bar s s \rangle \langle g_s^2 GG \rangle}{192\pi^2}-\frac{5 \alpha_s^2 m_s^4 \langle g_s^2 GG \rangle}{256\pi^4} \right) s \, ,
\\
\rho_{3^{+-}(j=2)}^{\bar{q}qgg}(s) &=& \frac{43\alpha_s^2 s^5}{26127360 \pi^4} + \left( \frac{5\alpha_s \langle g_s^2 GG \rangle }{62208 \pi^4} -\frac{25\alpha_s^2 \langle g_s^2 GG \rangle }{4644864 \pi^4}\right) s^3
\\ \nonumber &&+ \left( - \frac{ 7\alpha_s \langle g_s^3 G^3 \rangle}{41472\pi^3}- \frac{ \alpha_s^2  \langle \bar q q \rangle^2}{81}\right)s^2 +\frac{5 \langle g_s^2 GG \rangle^2 s}{41472\pi^2}\, ,
\\
\rho_{3^{+-}(j=2)}^{\bar{s}sgg}(s) &=& \frac{43\alpha_s^2 s^5}{26127360 \pi^4}- \frac{5 \alpha_s^2 m_s^2 s^4}{36288 \pi^4}  + \left( \frac{5\alpha_s \langle g_s^2 GG \rangle }{62208 \pi^4} -\frac{25\alpha_s^2 \langle g_s^2 GG \rangle }{4644864 \pi^4} + \frac{29 \alpha_s^2 m_s \langle \bar s s \rangle }{9072 \pi^2} + \frac{17\alpha_s^2 m_s^4 }{36288 \pi^4}\right) s^3
\\ \nonumber &&+ \left( - \frac{ 7\alpha_s \langle g_s^3 G^3 \rangle}{41472\pi^3}- \frac{ \alpha_s^2  \langle \bar s s \rangle^2}{81}-\frac{\alpha_s^2 m_s^3  \langle \bar s s \rangle}{648\pi^2} -\frac{ \alpha_s^2 m_s  \langle g_s \bar s \sigma G s \rangle}{216\pi^2}  +\frac{\alpha_s^2 m_s^2  \langle g_s^2 GG \rangle}{110592\pi^4}-\frac{\alpha_s m_s^2  \langle g_s^2 GG \rangle}{384\pi^3}\right) s^2
\\ \nonumber &&+ \left( \frac{5 \langle g_s^2 GG \rangle^2 }{41472\pi^2}-\frac{\alpha_s^2 m_s^2 \langle \bar s s \rangle^2}{81} +\frac{ \alpha_s m_s^2 \langle g_s^3 G^3 \rangle}{3456\pi^3}+\frac{5 \alpha_s m_s \langle \bar s s \rangle \langle g_s^2 GG \rangle}{324\pi}\right.
\\ \nonumber &&
\left.-\frac{65 \alpha_s^2 m_s \langle \bar s s \rangle \langle g_s^2 GG \rangle}{13824\pi^2}-\frac{5 \alpha_s^2 m_s^4 \langle g_s^2 GG \rangle}{6144\pi^4} \right) s \, ,
\\
\rho_{3^{--}(j=2)}^{\bar{q}qgg}(s) &=& \frac{\alpha_s^2 s^5}{653184 \pi^4} + \left( \frac{5\alpha_s \langle g_s^2 GG \rangle }{62208 \pi^3} +\frac{19\alpha_s^2 \langle g_s^2 GG \rangle }{387072 \pi^4}\right) s^3 + \left( - \frac{ \alpha_s \langle g_s^3 G^3 \rangle}{10368\pi^3}- \frac{ \alpha_s^2  \langle \bar q q \rangle^2}{27}\right)s^2
\\ \nonumber &&
+ \left( \frac{5 \langle g_s^2 GG \rangle^2 }{20736\pi^2}+\frac{10 \alpha_s^2 \langle \bar q q \rangle \langle g_s \bar q \sigma G q \rangle }{81}\right)s\, ,
\\
\rho_{3^{--}(j=2)}^{\bar{s}sgg}(s) &=& \frac{\alpha_s^2 s^5}{653184 \pi^4}- \frac{\alpha_s^2 m_s^2 s^4}{6048 \pi^4}  + \left( \frac{5\alpha_s \langle g_s^2 GG \rangle }{62208 \pi^3} +\frac{19\alpha_s^2 \langle g_s^2 GG \rangle }{387072 \pi^4} + \frac{37 \alpha_s^2 m_s \langle \bar s s \rangle }{6804 \pi^2} + \frac{\alpha_s^2 m_s^4 }{1008 \pi^4}\right) s^3
\\ \nonumber &&+ \left( - \frac{ \alpha_s \langle g_s^3 G^3 \rangle}{10368\pi^3}- \frac{ \alpha_s^2  \langle \bar s s \rangle^2}{27}-\frac{5\alpha_s^2 m_s^3  \langle \bar s s \rangle}{324\pi^2} -\frac{ \alpha_s^2 m_s  \langle g_s \bar s \sigma G s \rangle}{72\pi^2}  -\frac{11\alpha_s^2 m_s^2  \langle g_s^2 GG \rangle}{6912\pi^4}-\frac{\alpha_s m_s^2  \langle g_s^2 GG \rangle}{384\pi^3}\right) s^2
\\ \nonumber &&+ \left( \frac{5 \langle g_s^2 GG \rangle^2 }{20736\pi^2}+\frac{10 \alpha_s^2 \langle \bar s s \rangle \langle g_s \bar s \sigma G s \rangle }{81}+\frac{2\alpha_s^2 m_s^2 \langle \bar s s \rangle^2}{81} +\frac{ 13\alpha_s m_s^2 \langle g_s^3 G^3 \rangle}{3456\pi^3}+\frac{5 \alpha_s m_s \langle \bar s s \rangle \langle g_s^2 GG \rangle}{324\pi}\right.
\\ \nonumber &&
\left.+\frac{73 \alpha_s^2 m_s \langle \bar s s \rangle \langle g_s^2 GG \rangle}{6912\pi^2}+\frac{ \alpha_s^2 m_s^4 \langle g_s^2 GG \rangle}{3072\pi^4} \right) s \, ,
\\
\rho^{\bar q q gg}_{3^{-+}(j=2)}(s) &=& \frac{\alpha_s^2 s^5}{115200 \pi^4}+\left( \frac{\alpha_s^2 \langle g_s^2 GG \rangle }{6912 \pi^3}+\frac{29\alpha_s^2 \langle g_s^2 GG \rangle }{331776 \pi^4}\right) s^3 +\left(
- \frac{ \alpha_s^2 \langle \bar s s \rangle^2}{9}
- \frac{  \alpha_s \langle g_s^3 G^3 \rangle}{3840\pi^3}\right) s^2
\\ \nonumber &&
+  \frac{2\alpha_s^2 \langle \bar s s \rangle \langle g_s \bar s \sigma G s \rangle s}{9}   \, ,
\\
\rho^{\bar s s gg}_{3^{-+}(j=2)}(s) &=& \frac{\alpha_s^2 s^5}{115200 \pi^4}- \frac{\alpha_s^2 m_s^2 s^4}{1260 \pi^4} +
\left( \frac{\alpha_s^2 \langle g_s^2 GG \rangle }{6912 \pi^3}
+\frac{29\alpha_s^2 \langle g_s^2 GG \rangle }{331776 \pi^4}
+ \frac{\alpha_s^2 m_s^4 }{288 \pi^4}
+ \frac{23\alpha_s^2 m_s \langle \bar s s \rangle }{1080 \pi^2}\right) s^3
\\ \nonumber &&+\left(- \frac{ \alpha_s^2 \langle \bar s s \rangle^2}{9}
- \frac{  \alpha_s \langle g_s^3 G^3 \rangle}{3840\pi^3}
-\frac{ \alpha_s^2 m_s  \langle g_s \bar s \sigma G s \rangle}{24\pi^2}
-\frac{ \alpha_s^2 m_s^3  \langle \bar s s \rangle}{30\pi^2}
-\frac{3 \alpha_s m_s^2 \langle g_s^2 GG \rangle}{640\pi^3}
-\frac{27 \alpha_s^2 m_s^2 \langle g_s^2 GG \rangle}{10240\pi^4}
\right) s^2
\\ \nonumber &&+
\left(  \frac{2\alpha_s^2 \langle \bar s s \rangle \langle g_s \bar s \sigma G s \rangle}{9}
+ \frac{\alpha_s m_s^2 \langle g_s^3 G^3 \rangle}{384\pi^3}
+\frac{ \alpha_s m_s \langle g_s^2 GG \rangle \langle \bar s s \rangle}{36\pi}
+\frac{ \alpha_s^2 m_s \langle g_s^2 GG \rangle \langle \bar s s \rangle}{64\pi^2}
\right) s \, .
\end{eqnarray}
\end{widetext}

\bibliographystyle{elsarticle-num}
\bibliography{ref}

\end{document}